\documentclass[11pt]{article}
\usepackage{epsfig, amsmath, amssymb, amsthm, times,color}

\usepackage[utf8]{inputenc}

\usepackage{amsmath,amssymb,amsmath}
\usepackage{mathabx}
\usepackage{graphicx}
\usepackage{subfigure}
\usepackage{cite}
\usepackage{hyperref}
\usepackage{url}
\usepackage{mathrsfs}

\numberwithin{equation}{section}

\newcommand{\R}{\mathbb{R}}
\newcommand{\T}{\mathbb{T}}
\newcommand{\C}{\mathbb{C}}
\newcommand{\Z}{\mathbb{Z}}
\newcommand{\N}{\mathbb{N}}

\newcommand{\bW}{\boldsymbol{W}}
\newcommand{\bL}{\boldsymbol{L}}
\newcommand{\bB}{\boldsymbol{B}}

\newcommand{\bS}{\boldsymbol{S}}

\newcommand{\cH}{\mathcal{H}}

\newcommand{\cC}{\mathcal{C}}

\newcommand{\bbH}{\mathbb{H}}

\def\lbl{\label}
\def\be{\begin{equation}}
	\def\ee{\end{equation}}

\newcommand{\p}{{\partial}}
\newcommand{\iu}{{i\mkern1mu}}

\def\1{\mathbf{1}}

\newtheorem{theorem}{Theorem}[section]

\newtheorem{lemma}[theorem]{Lemma}

\newtheorem{remark}[theorem]{Remark}

\newtheorem{assumption}[theorem]{Assumption}

\title{Bifurcations and patterns in the Kuramoto model with inertia}
\date{\today}
\author{Hayato Chiba\thanks{Advanced Institute for Materials Research,
		Tohoku  University, Sendai, 980-8557, Japan, {\tt hchiba@tohoku.ac.jp}},\;
	Georgi S. Medvedev\thanks{Department of Mathematics, 
		Drexel University, 3141 Chestnut Street, Philadelphia, PA 19104,
		{\tt medvedev@drexel.edu}},
	\;
	and
	\;
	Matthew S. Mizuhara\thanks{Department of 
		Mathematics and Statistics,
		The College of New Jersey,
		{\tt  mizuharm@tcnj.edu}}
}

\begin{document}
	\maketitle
	\begin{abstract}
          In this work, we analyze the Kuramoto model (KM) with inertia  on a convergent family of graphs.
          It is assumed that the intrinsic frequencies of the individual oscillators
          are sampled from a probability
          distribution. In addition, a given graph, which may also be random, assigns
          network connectivity. As in the original KM, in the model with inertia, the weak
          coupling regime features mixing, the state of the network when the phases
          (but not velocities) of all oscillators are distributed uniformly around the unit circle.
          We study patterns, which emerge when mixing loses stability under the variation
          of the strength of coupling. We identify a pitchfork (PF) and an Andronov-Hopf (AH)
          bifurcations in the model with multimodal intrinsic frequency distributions.
          To this effect, we use a combination of the linear stability analysis
          and Penrose diagrams, a geometric technique for studying stability of mixing.	
          We show that the type of a bifurcation and a nascent spatiotemporal pattern depend
          on the interplay of the qualitative properties of the intrinsic
          frequency distribution and 
          network connectivity. 	
	\end{abstract}
	
	
\section{Introduction}
\setcounter{equation}{0}

In this paper, we study the following system of coupled second order damped oscillators on a
convergent sequence of
graphs $\{\Gamma_n\}$:
\begin{equation}\label{2KM}
  \ddot\theta _i+ \gamma  \dot\theta _i= \omega _i +
  \frac{2K}{n} \sum^n_{j=1}a^n_{ij} \sin \left(\theta _j-\theta _i\right),
\quad i\in [n]:=\{1,2,\dots,n\},
\end{equation}
where $\theta _i: \R^+\to\T:=\R/2\pi\Z$ denotes the phase of the $i$th oscillator,
$\gamma >0$ is a damping constant, $K$ is the coupling strength and
$a^n_{ij}$ is the adjacency matrix of $\Gamma_n$. Intrinsic frequencies $\omega_i, i\in [n],$
are independent identically distributed  random variables drawn from the probability
       distribution with density $g$.
       By rescaling time, intrinsic frequencies, and $K$, one can make $\gamma=1$,
       which will be assumed without loss of generality throughout this paper.
\begin{figure}
		\centering
		\begin{minipage}[b]{.98\textwidth}
			\textbf{a}~\includegraphics[width=0.22\textwidth]{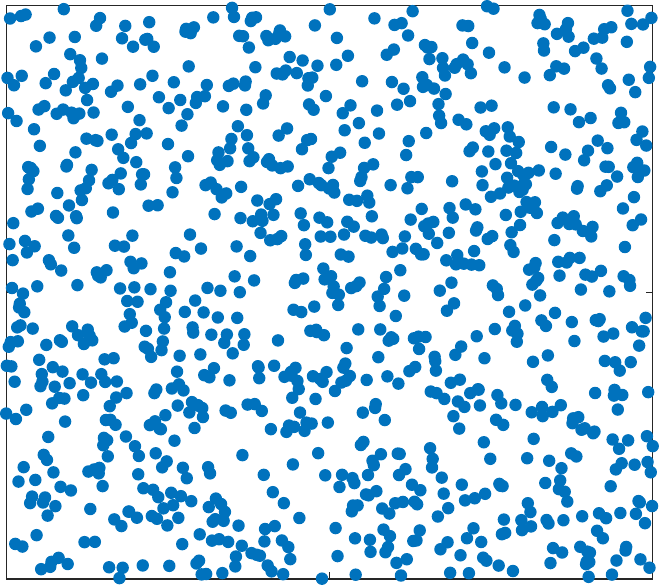}
			\hfill
			\textbf{b}~\includegraphics[width=0.22\textwidth]{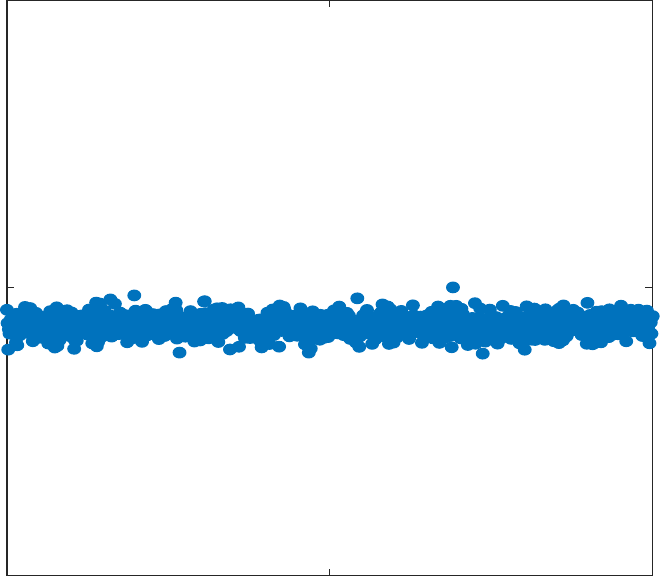}
			\hfill
			\textbf{c}~\includegraphics[width=0.22\textwidth]{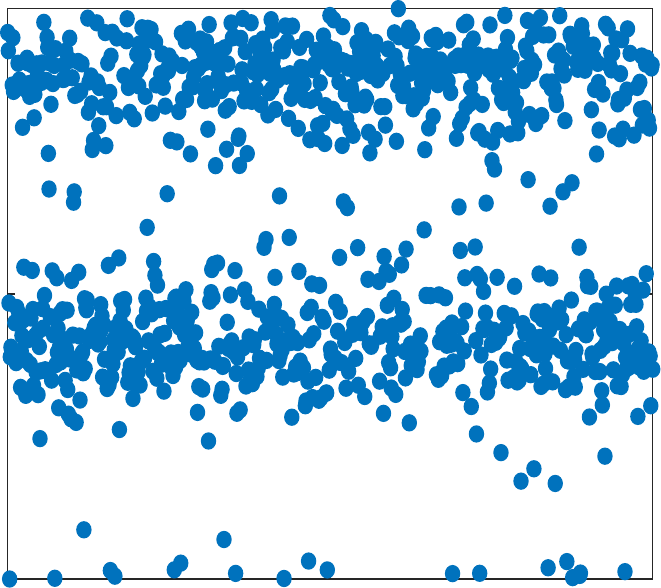}
			\hfill
			\textbf{d}~\includegraphics[width=0.22\textwidth]{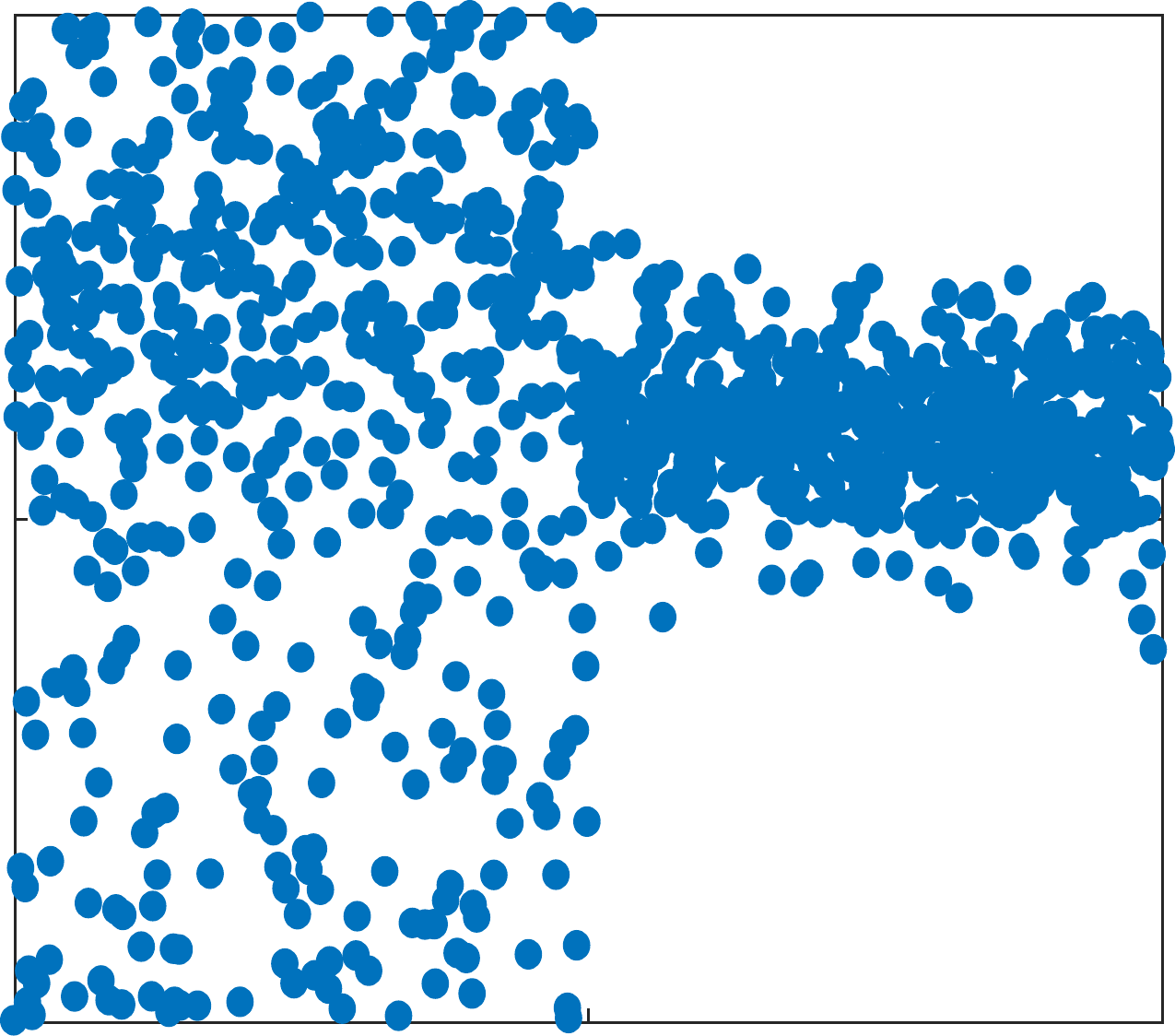}\\
                           \\
			\textbf{e}~\includegraphics[width=0.22\textwidth]{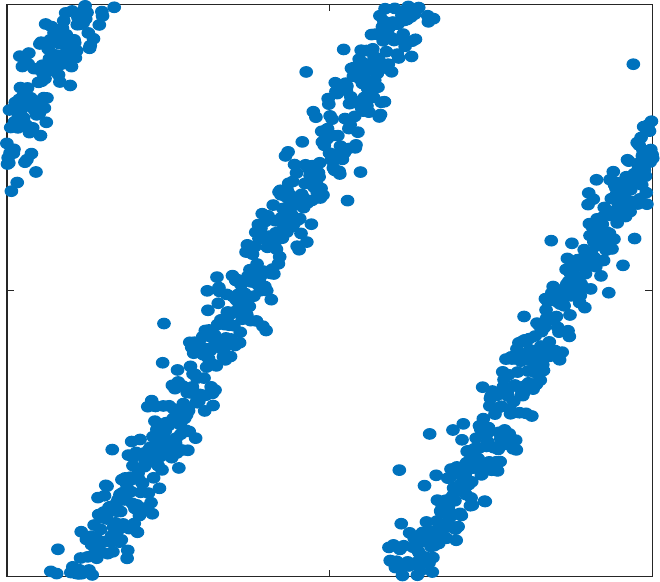}
			\textbf{f}~\includegraphics[width=0.22\textwidth]{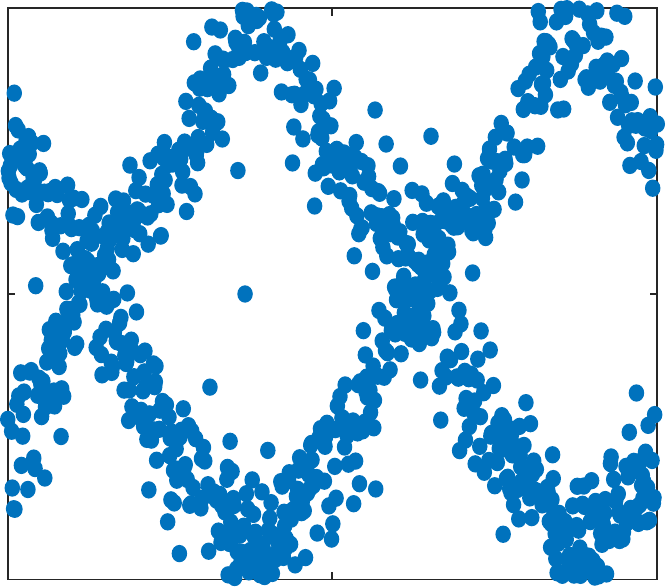}
			\hfill
			\textbf{g}~\includegraphics[width=0.22\textwidth]{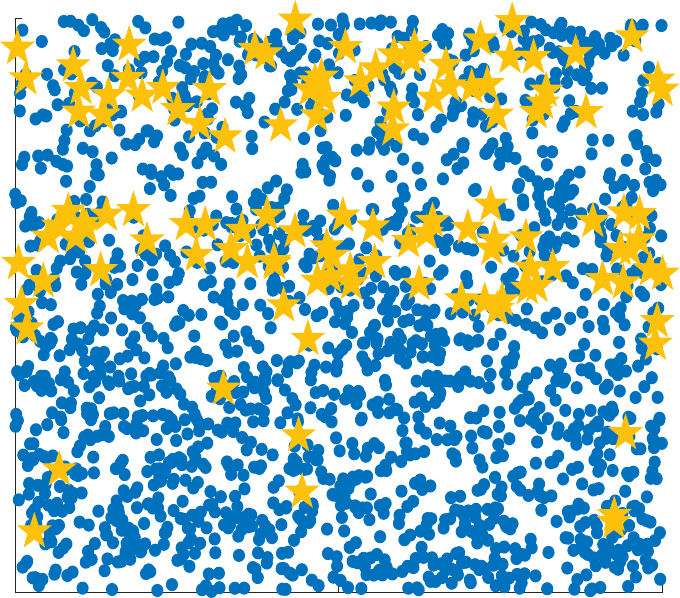}
			\hfill
			\textbf{h}~\includegraphics[width=0.22\textwidth]{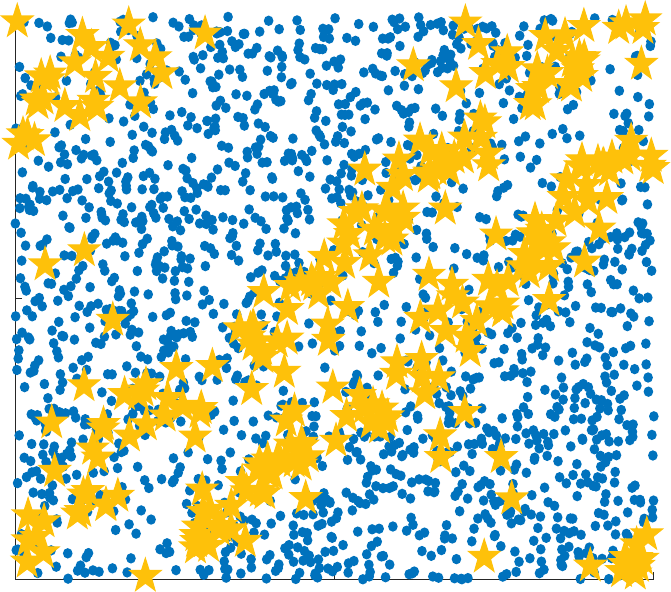}
			\hfill
			\caption{A collection of patterns generated by \eqref{2KM}
                          with unimodal and bimodal frequency distributions on complete and nonlocal nearest--neighbor
                          graphs: a) mixing, b) synchronization, c) clusters, d) chimeras, e) twisted states, f)
                          two sets of twisted states traveling in opposite directions,
                          g) a two-cluster partially locked state, h) a  two-cluster twisted partially locked state.
                          In g) and h) yellow stars correspond stationary clusters superimposed onto  irregularly
                          moving oscillators. Patterns in a)-f) can also be generated using a classical (first order)
                          Kuramoto model using the same settings (cf.~\cite{MM22}). Patterns in g) and h) are
                          new.
                          {Unless otherwise stated,  all simulations of \eqref{2KM} were
                            performed with $n=10^3$  and $\gamma = 1$. The snapshots \textbf{b}-\textbf{h}
                            are representative for the dynamical states bifurcating from the incoherent state shown in
                            \textbf{a}. The range of $K$ and the network connectivity corresponding to these patterns are explained
                            in the text and in the bifurcation diagrams below.}                            
			}\label{f.patterns}
		\end{minipage}
              \end{figure}
              
       The system of equations \eqref{2KM} is a generalization of the KM of coupled phase oscillators,
       which provides an important framework for studying collective dynamics in coupled networks
       \cite{Kur75, Rod16}.
       This is also an established model of a power network \cite{TOS19, DB12, SMV84}.
       The second order derivatives are used to incorporate inertial effects into the system's dynamics.
       Thus, the name - the KM with inertia. Compared to the original KM, it is a more
       flexible model with more degrees of freedom and richer dynamics. It features a range of
       spatiotemporal patterns found in the original KM \cite{CMM20, MM22} as well as a few new
       ones that have not been studied before (Figure~\ref{f.patterns}).

       The KM is best known for the phase transition from a highly irregular mixing behavior in the weak coupling
       regime to a gradual buildup of coherence leading to synchronization \cite{Kur75, StrMir91}.
       In the first order KM with intrinsic frequencies sampled from a symmetric unimodal distribution,
       the transition to synchrony lies through the pitchfork (PF) bifurcation of mixing \cite{Chi15a, Die16}. A similar mechanism
       is involved in the transition to synchronization in the KM on graphs \cite{ChiMed19a, CMM18}. In the KM
       with more general coupling functions or with multimodal frequency distribution mixing may lose stability
       through the Andronov-Hopf (AH) bifurcation \cite{Chi21, CMM20, MM22}. Qualitative properties of the intrinsic frequency
       distribution and the properties of network topology  translate into a variety of spatiotemporal patterns
       replacing mixing when it loses stability \cite{CMM20, MM22}. Following \cite{CMM20, MM22}, we use a
       combination of the linear stability analysis \cite{ChiMed22} and Penrose diagrams
       \cite{Penrose60, Die16, MM22}
       to describe bifurcations leading to the loss of stability of mixing as well as emerging patterns in the KM with
       inertia. We find many parallels in pattern formation mechanisms involved in both the ordinary KM and
       the KM with inertia, but we
       also find patterns that are not present in the former model. For instance, we identified a class of
       partially locked states (PLS) based on stationary antiphase clusters (cf.~\cite{MM21})
       (see Fig.~\ref{f.patterns} \textbf{g}, \textbf{h}). We show that
       the KM with inertia with symmetric bimodal (with well separated peaks) frequency distribution  and nonlocal
       nearest neighbor coupling has four distinct bifurcation
       scenarios (cf.~\eqref{K-bi}), compared to a single one for
       the ordinary KM in a similar setting (see Section~\ref{sec.connect}).
       Overall, the KM with inertia offers a richer repertoire of patterns.
       
{
       The loss of stability of the incoherent state in the KM with inertia on with all-to-all coupling was studied
       in \cite{TanLichOish97,  JiPerRod2104, AceBonSpi2000, GupCamRuf2014}. The results in these papers
       are based on self-consistent analysis and numerical simulations. Our approach relies on a rigorous
       linear stability analysis and applies to the KM with multimodal frequency distributions on convergent
       graph sequences. We show that the loss of stability of the incoherent state may lead to a variety of
       interesting spatiotemporal patterns, which can be related to the qualitative properties of the intrinsic
       frequency distribution and network topology.
     }
     
       The paper is organized as follows. The next section presents necessary information about stability of mixing
       adapted from \cite{ChiMed22}. In Section~\ref{sec.Penrose}, we analyze the loss of stability of mixing in the
       model with a unimodal intrinsic frequency distribution. We use this problem as a convenient setting to introduce
       Penrose diagrams. The full power of this technique is revealed when we apply it to study bifurcations in the model
       with a family of bimodal intrinsic frequency distributions in Section~\ref{sec.bimod}.
       Here, in addition, to the PF bifurcation of mixing which
       already appeared in the unimodal case, we encounter an AH bifurcation. The latter is responsible for formation
       of traveling clusters. Furthermore, we describe another PF bifurcation, which occurs at a negative value of $K$.
       In contrast to the PF in the model with unimodal distribution, this time the PF bifurcation supports  stationary
       antiphase clusters superimposed onto irregularly moving oscillators. This pattern was not present in
       the ordinary KM in (cf~\cite{MM22}). In Section~\ref{sec.connect}, we describe the role of the network topology
       on patterns emerging when mixing loses stability.
       We end this paper with concluding remarks in Section~\ref{sec.discuss}.

       \section{The linear stability of mixing}\label{sec.stability}
       \subsection{The mean field limit}
       The first step in the analysis of mixing is the derivation of the mean field limit for \eqref{2KM}. To this
       end, we rewrite \eqref{2KM} as follows
\be\lbl{re2KM}
\begin{split}
\dot\theta _i &  = \psi_i + \omega _i, \\
\dot\psi_i  &= -\psi_i + \frac{2K}{n} \sum^n_{j=1}a^n_{ij} \sin (\theta _j-\theta _i),\\
\dot\omega_i& =0,\qquad i\in [n].
\end{split}
\ee
Equation \eqref{re2KM} shows that the phase space of each (uncoupled) oscillator is $\T\times\R\times\R$.
With \eqref{re2KM} at hand, it is straightforward to write down the Vlasov equation  describing the dynamics of
\eqref{re2KM} in the limit as $n\to\infty$ (cf. \cite{Gol16})
\begin{equation}\label{Vlasov}
\partial_t f + \partial_\theta \left( (\psi + \omega ) f \right)
+\partial_\psi \left( (-\psi + \mathbf{N}[f])f \right) = 0,
\end{equation}
where
\begin{equation*}
\displaystyle \mathbf{N}[f](t, \theta ,x) = \frac{K}{i}\left( e^{-i\theta } h(t,x) - e^{i\theta } \overline{h(t,x)} \right)
\end{equation*}
and
\begin{equation}\lbl{def-order}
h(t,x) = \int_{\T\times\R^2 \times I} W(x,y) e^{i \theta } f(t,\theta ,\psi, \omega ,y) d\theta d\psi d\omega dy,
\end{equation}
is the local order parameter. Here, $I = [0,1]$ and $f(t,\theta,\psi,\omega,x) d\theta d\psi d\omega$ stands for the
probability that the state of the oscillator at the `spatial' location $x\in I$ and time $t\ge 0$ is in
$[\theta, \theta+d\theta)\times [\psi, \psi+d\psi)\times [\omega, \omega+d\omega)$.
$W$ is a square integrable function on $I^2$, which describes the limit of the
graph sequence $\{\Gamma_n\}$\footnote{Such functions are called graphons in the graph theory.
For the details on graphons, graph limits, and their applications to the continuum description of
the dynamical networks, we refer the interested
reader to \cite{ChiMed19a, Med19}.}.

 A distribution--valued solution $f(t,\theta,\psi,\omega,x)$ of the initial value problem for the Vlasov equation \eqref{Vlasov}
yields the probability distribution of particles in the phase space $\T\times \R^2$ for each
$(t,x)\in \R^+\times I$ (cf. \cite{Dob79, KVMed18}).  By mixing we mean the following
steady state solution of \eqref{Vlasov}
\be\lbl{mixing}
f_{mix}=\frac{g(\omega)}{2\pi}\delta(\psi),
\ee
where $\delta$ stands for Dirac's delta function. $f_{mix}$ corresponds to the stationary regime, which is
characterized by the uniform distribution of the phases $\theta_i, i\in [n]$.

\subsection{The linearized equation}\lbl{sec.Fourier}
In this section, we present the key
steps in the stability analysis and refer the reader to \cite{ChiMed22} for more details.

Throughout this paper, we will assume the following.
\begin{assumption}\label{as.decay}
  Let $g$ be a real analytic function. In addition, we assume that
  $\hat{g}(\eta) :=\int_\R e^{\iu \eta\omega}g(\omega) d\omega$ is a
  continuous function such that
  \be\lbl{g-decay}
  \lim_{\eta\to\infty} |\hat{g}(\eta)|e^{a\eta}=0
  \ee
  for some $0<a<1$.
  \end{assumption}
  \begin{remark}
By the Paley-Wiener theorem, \eqref{g-decay} implies that    
$g(\omega )$ has an analytic continuation to the region $0\le \mathrm{Im}(z) < a$. 
\end{remark}

Since $\omega$ does not change in time, $f$ satisfies the following constraint
\begin{eqnarray}\label{g-constraint}
g(\omega ) = \int_{\T \times \R}\!  f(t,\theta ,\psi, \omega ,x)
  d\theta d\psi, \quad \forall (t,x)\in \R^+\times I.
\end{eqnarray}

Next, we recast \eqref{Vlasov} in Fourier variables
\begin{equation}\label{Fourier}
\p_tu_j = 
(j-\zeta) \partial_\zeta u_j + j \partial_\eta u_j
     + K\zeta (h(t,x) u_{j-1} - \overline{h(t,x)} u_{j+1}),\; j\in\Z,
\end{equation}
where we used
\begin{equation}\label{transform}
u_j(t, \zeta, \eta , x) = \int_{\R^2\times \T} e^{i(j\theta + \zeta \psi + \eta \omega  )} 
   f (t,\theta ,\psi, \omega ,x) d\theta d\psi d\omega,\quad j\in\Z
\end{equation}
and integration by parts. Note that the expression for the
local parameter can be rewritten as
\begin{equation*}
h(t,x) = \int_I W(x,y) u_1(t,0,0 ,y) dy.
\end{equation*}
In addition, we have the following constraints
\begin{eqnarray}\label{F-const-a}
  u_0(t,0,0 ,x) &=& 1,\\
  \label{F-const-b}
  u_{-j}(t,-\zeta, -\eta ,x) & =& \overline{u_j (t,\zeta,\eta ,x)}, \; j\in \N.
\end{eqnarray}
Equation  \eqref{F-const-a} follows from \eqref{g-constraint}. Equation 
\eqref{F-const-b} follows from the fact that $f$ is real. Thus,
it is sufficient to restrict to $j\in \N\bigcup\{0\}$ in \eqref{Fourier}.

By changing from $\zeta$ to $\xi$ given by the following relations:
\begin{equation}
\left\{ \begin{array}{ll}
\zeta - j = - e^{-\xi_j}, & \zeta -j<0, \\
\zeta - j =   e^{-\xi_j}, & \zeta -j\geq 0, \\
\end{array} \right.
\label{1-6}
\end{equation}
and setting
\begin{equation}
v_j (t, \xi_j, \eta ,x) := 
\left\{ \begin{array}{ll}
u_j(t, j- e^{-\xi_j }, \eta , x), & \zeta -j<0, \\[0.2cm]
u_j(t, j+ e^{-\xi_j }, \eta , x), & \zeta -j\geq 0, \\
\end{array} \right.
\end{equation}
we obtain
\begin{eqnarray}\label{rewrite-v}
\partial_t v_j &
  = &\partial_{\xi_j} v_j 
      +j \partial_\eta v_j+K(j-e^{-\xi_j}) \left( h(t,x)v_{j-1}-\overline{h(t,x)} v_{j+1}\right),\quad j\in \Z\\
  \label{rewrite-h}  
\displaystyle h(t,x)& = &\int_I  W(x,y) v_1(t,0,0,y)dy,                
\end{eqnarray}
subject to the constraint $\lim_{\xi\to\infty} v_0(t, \xi , \eta ,x) = \hat{g}(\eta)$.
For $j\geq 0$, we adopt the first line of (\ref{1-6}) because in the definition of
the local order parameter, we need $u_1(t,0,0,x)$, for which $\zeta -j = -1<0$.
By the same reason, we use the second line for $j \leq -1$.


The steady state of the Vlasov equation, $f_{mix}$,  in the Fourier space has the following form
\be\lbl{Fmix}
v_0 = \hat{g}(\eta), \quad v_j = 0, j\in\N.
\ee

To investigate the stability of \eqref{Fmix}, let $w_0 = v_0 - \hat{g}(\eta)$ and $w_j = v_j$ for $j\neq 0$.
Then we obtain the system
\begin{equation}
\left\{ \begin{array}{l}
\displaystyle \partial_t w_1= \partial_{\xi_1} w_1  + \partial_\eta w_1
     + K(1- e^{-\xi_1}) \left( h(t,x) \hat{g}(\eta) + h(t,x)w_0 - \overline{h(t,x)} w_2 \right), \\[0.4cm]
\partial_t w_j = \partial_{\xi_j} w_j  + j\partial_{\eta} w_j
          + K(j- e^{-\xi_j}) \left( h(t,x)w_{j-1} - \overline{h(t,x)} w_{j+1} \right), \; j\ge 0 \,\, \text{and} \,\,  j\neq 1.\\
          [0.4cm]
\displaystyle h(t,x) = \int_I W(x,y) w_1(t,0 ,0 ,y) dy,
\end{array} \right.
\label{1-8}
\end{equation}
and $\lim_{\xi\to\infty} w_0(t, \xi, \eta ,x) = 0$. 
Our goal is to investigate the stability and bifurcations of the steady state (mixing)
$w_j = 0, \, j\in \Z$ of this system.

The linearized system has the following form
\begin{eqnarray}\label{linear}
\partial_t w_1 & = & \bL_1[w_1] + K \bB[w_1] =: \bS[w_1],  \\
\partial_t w_j &=& \bL_j [w_j],\quad \;j\ge 0 \,\, \text{and} \,\,  j\neq 1, \label{linear-2}
\end{eqnarray}
where
\begin{eqnarray} \lbl{Lj}
 \bL_j[\phi](\xi, \eta , x) 
  &=& \left( {\partial_\xi} + j {\partial_\eta}\right) \phi(\xi, \eta, x), \; j\in \Z\\
  \label{def-B}
\bB[\phi](\xi, \eta , x)&=&(1-e^{-\xi}) \hat{g}(\eta) \bW[\phi (0,0, \cdot)](x),
\end{eqnarray}
and
\begin{equation}\lbl{def-W}
\bW[f](x) = \int_{\R} W(x,y) f(y) dy.
\end{equation}
The self-adjoint operator $\mathbf{W}$ reflects the impact of connectivity
on stability of mixing.

\subsection{The spaces}
To proceed with the linear stability analysis of the mixing state, we need
to introduce the following Banach spaces.

Recall $a\in (0,1)$ defined in Assumption~\ref{as.decay}. For $\alpha\in \{0,1\}$, let
\begin{equation}\label{weights}
  \beta^+_1(\eta) = \max\{ 1, e^{\alpha\eta}\}, \quad \beta^-_1(\eta) = \min\{ 1, e^{\alpha\eta}\},
  \quad\mbox{and}\quad \beta_2(\xi) = \min\{ e^{\xi}, 1 \},
  \end{equation}
and define
\begin{equation}\label{spaces}
\begin{split}
  \mathcal{X}^\pm_\alpha  & = \{ \phi : \text{continuous on $\R$},\, 
       \| \phi \|_{\mathcal{X}^\pm_\alpha}   = \sup_{\eta} \beta^\pm_1(\eta) |\phi (\eta)| < \infty \}, \\
\mathcal{Y}^\pm_\alpha  &= \{ \phi : \text{continuous on $\R^2$},\, 
       \| \phi \|_{\mathcal{Y}^\pm_\alpha} = \sup_{\xi, \eta} \beta^\pm_1(\eta)\beta_2(\xi) |\phi (\xi, \eta)| < \infty \}, \\
       \mathcal{H}^\pm_\alpha  &= L^2 (I; \mathcal{Y}^\pm_\alpha).
     \end{split}
\end{equation}
{Recall that $I$ stands for $[0,1]$.}
The norms on $\mathcal{H}^{\pm}_\alpha$ are defined by
\begin{equation}\label{norm-Ha}
  \| \phi \|^2_{\mathcal{H}^{\pm}_\alpha}=\int_{I}\! \left( \sup_{\xi, \eta}  \beta^{\pm}_1(\eta)
    \beta_2(\xi) |\phi (\xi, \eta, x)|\right)^2 dx.
\end{equation}
Note that the spaces
$$
\mathcal{H}^+_a  \subset \mathcal{H}^+_0 = \mathcal{H}^-_0 \subset \mathcal{H}^-_a
$$
form a Gelfand triplet.
Since the linear operators defined above have essential spectra on the imaginary axis,
we need the generalized spectral theory based on the Gelfand triplet to detect bifurcations of mixing \cite{ChiMed22}.
 

\subsection{The spectrum of $\bS$}\label{sec.spectrum}
Recall the definition of $\bW$ \eqref{def-W} and note that it is a compact self-adjoint 
operator on $L^2(I)$. Therefore, the eigenvalues of $\bW$ are real with the
only accumulation point at $0$. We denote the set of eigenvalues of $\bW$ by
$\sigma _p(\bW)$.

Operators  $\bL_j, j\in\Z$ and $\bB$ are densely defined on $\mathcal{H}^+_0$ (see \eqref{Lj},
\eqref{def-B}) and $\bB$ is a bounded operator (cf. \cite{ChiMed22}).
 The resolvent of $\bL_j$ is given by
 \begin{equation}\label{rewrite-res}
(\lambda - \bL_j)^{-1}[v](\xi, \eta, x ) = \left\{ \begin{array}{ll}
\displaystyle \int^{\infty}_{0}\! e^{-\lambda t} v (\xi + t, \eta + jt,x)dt, & \;\Re~\lambda  >0,\\
\displaystyle  -\int^{0}_{-\infty} \! e^{-\lambda t} v (\xi + t, \eta + jt,x)dt & \;\Re~\lambda <0.
                                                          \end{array}
                                                        \right.
\end{equation}
The right--hand side of \eqref{rewrite-res} belongs to $\mathcal{H}^+_0$ for any
$v\in \mathcal{H}^+_0$ only if $\Re~\lambda < -1$ or $\Re~\lambda>0$.
For $ -1\leq  \Re~\lambda\ \leq 0$, the set of $v$ such that the right-hand side exists is not dense
in $\mathcal{H}^+_0$. Thus, the residual spectrum of $\bL_j$ is the region
$\mathbb{S}_1:=\{ z\in\C:\; -1 \le \Re z\le 0 \}$. $\mathbb{S}_1$ contains no eigenvalues of $\bL_j$, $j\in\Z$.
The essential spectrum of $\bS$ is  given by $\mathbb{S}_1$,
because $\bS = \bL_1 + K\bB$ is the bounded perturbation of
$\bL_1$. However, $\mathbb{S}_1$ may still contain eigenvalues of $\bS$,
as one can see from the following lemma.

\begin{lemma} (cf.~\cite{ChiMed22})
  \label{lem.eig-S}
  Let $\nu$ be a nonzero eigenvalue of $\bW$ and let $V_\nu\in L^2(I)$ be a corresponding
  eigenfunction. Define
  \be\lbl{define-D}
D(\lambda, \xi, \eta)=\int_{\R} \left( \frac{1}{\lambda -\iu\omega } -
        \frac{e^{-\xi}}{\lambda +1-\iu\omega } \right)
      e^{\iu\eta \omega } g(\omega)d\omega,
      \ee
and
  \begin{equation}\label{def-D}
    G(\lambda):=D(\lambda, 0,0) = 
 \int_{\R} \left( \frac{1}{\lambda -\iu\omega } -
  \frac{1}{\lambda +1-\iu\omega } \right) g(\omega )d\omega.
\end{equation}
Then the root $\lambda=\lambda(\nu)$ of 
the following equation 
\begin{equation}\label{Deqn}
G(\lambda ) = \frac{1}{K\nu},
\end{equation}
not belonging to $\p\sigma(\bL_1)=\{z\in\C: \Re~z=-1 \;\mbox{or}\; \Re~z=0\}$,
is an eigenvalue of $\bS$ on $\cH^+_0$.
For each such root $\lambda=\lambda(\nu)$
the corresponding eigenfunction is given by
\begin{equation} \label{eigen-fun}
v_\lambda(\xi, \eta ,x) 
= D(\lambda, \xi, \eta) V_\nu(x).
\end{equation}
\end{lemma}

To study patterns arising at the bifurcations of mixing we define the function 
  \begin{equation}\label{def-Ups}
\Upsilon_\lambda(\omega,\xi)  =\left( \frac{1}{\lambda -\iu\omega } -
  \frac{e^{-\xi}}{\lambda +1-\iu\omega } \right) g(\omega).
\end{equation}
Then, the following equality holds:
  \begin{equation}\label{w-lambda}
v_\lambda(\xi,\eta,x) = {F}_\eta[\Upsilon_\lambda](\xi,\eta) V_\nu (x),
\end{equation}
where $F_\eta$ is the Fourier transform with respect to $\omega \mapsto \eta$.

Let
\begin{equation}\label{def-Ha}
  \mathbb{H}_a:=\{z\in\mathbb{C}:~\Re z>-a\}.
\end{equation}
For $\Re~\lambda  >0,$ $\Upsilon_\lambda(\omega,\xi)$ is
an integrable function in $\omega$ as can be seen from \eqref{def-Ups}.
For $\lambda=\iu u\in \iu\R$,  $\Upsilon_\lambda(\omega,\xi)$ is
no longer an integrable function, but  
it can be interpreted as a tempered distribution,
as the following argument shows. 
By Sokhotski--Plemelj formula (cf.~\cite{Simon-Complex}), we have
           \begin{alignat*}{2}
           \lim_{x \to 0+} \langle  \Upsilon_{x+\iu y},\phi\rangle & =\lim_{x \to 0+}
           \int_{-\infty}^\infty {g(\omega)\phi(\omega)\over x+\iu y-\iu\omega}d\omega
-\int_{-\infty}^\infty \frac{e^{-\xi} g(\omega)\phi(\omega)}{1+\iu (y-\omega)}d\omega\\
           \\
&=\pi g(y)\phi(y)+\iu\, \operatorname{p.\!v.}
\int_{-\infty}^\infty {g(\omega+y)\phi(\omega+y)\over \omega} d\omega
-\int_{-\infty}^\infty \frac{e^{-\xi} g(\omega)\phi(\omega)}{1+\iu (y-\omega)}d\omega,
\end{alignat*}
for any $\phi$ from the Schwartz class $\mathcal{S}(\R)$.
Thus, $\Upsilon_{\iu y}:=\lim_{x\to 0+} \Upsilon_{x+iy}\in\mathcal{S}^\prime (\R)$ for each $\xi$ and 
          \be\lbl{Ups-iy}
    \Upsilon_{\iu y} =
    \pi g(y) \delta_{y}+\iu\mathcal{P}_{y}[g]
    - \frac{e^{-\xi} g(\omega)}{1+\iu (y-\omega)},
           \ee
           where $\delta_y$ stands for the Dirac's delta function supported at $y$ and
           \begin{equation}\label{pv-distr}
           \langle \mathcal{P}_y[g], \phi \rangle=\operatorname{p.\!v.}
             \int_{-\infty}^\infty {g(\omega+y)\phi(\omega+y)\over \omega}d\omega.
           \end{equation}
           Indeed, $\lim_{\lambda \to 0+} v_{\lambda} (\xi, \eta, x)$ is an element of the space $\mathcal{H}^{-}_a$,
           the dual space of $\mathcal{H}^{+}_a$.
This fact was used to apply the center manifold reduction on $\mathcal{H}^{-}_a$ in \cite{ChiMed22}.
    

           The formulae for eigenfunctions of $\bS$ \eqref{w-lambda}-\eqref{Ups-iy} corresponding to
           bifurcating eigenvalues will be used below to explain spatiotemporal  patterns arising at the
           loss of stability of mixing.

	\section{The method of Penrose}\label{sec.Penrose}
	\setcounter{equation}{0}

Having reviewed the linear stability analysis, we next focus on 
the instability of mixing. To this end, we use a geometric 
method for locating and identifying bifurcations in the 
Vlasov equation, which was invented by
Penrose in the context of Landau damping \cite{Penrose60}. This method
was adapted to the analysis of the classical Kuramoto model in
\cite{Die16} and  the Kuramoto model on graphs in \cite{CMM20, MM22}.

\begin{figure}
		\begin{center}
			\textbf{a}\includegraphics[width=0.45\textwidth]{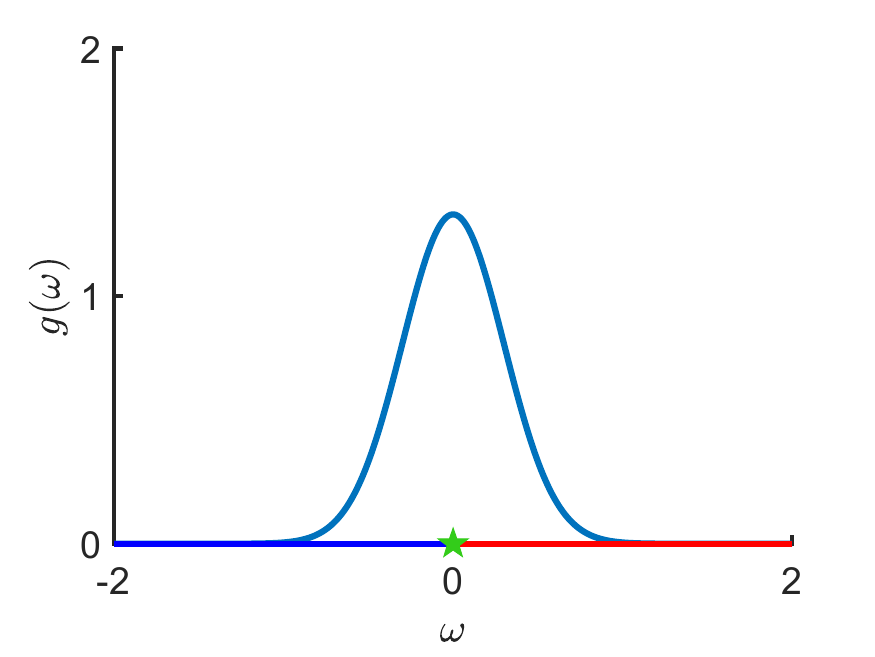}
			\textbf{b}\includegraphics[width=0.45\textwidth]{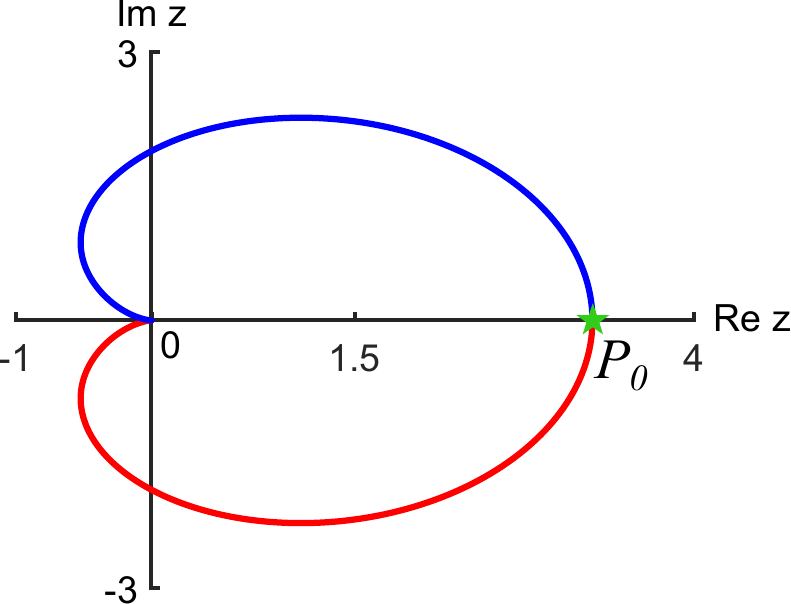}\\
			\textbf{c}\includegraphics[width=0.45\textwidth]{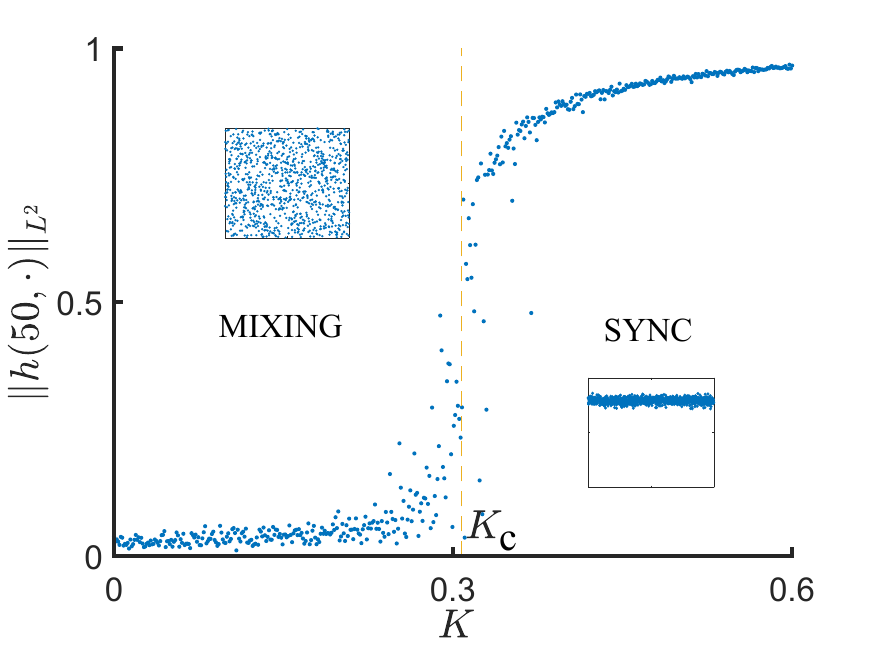}
			\textbf{d}\includegraphics[width=0.45\textwidth]{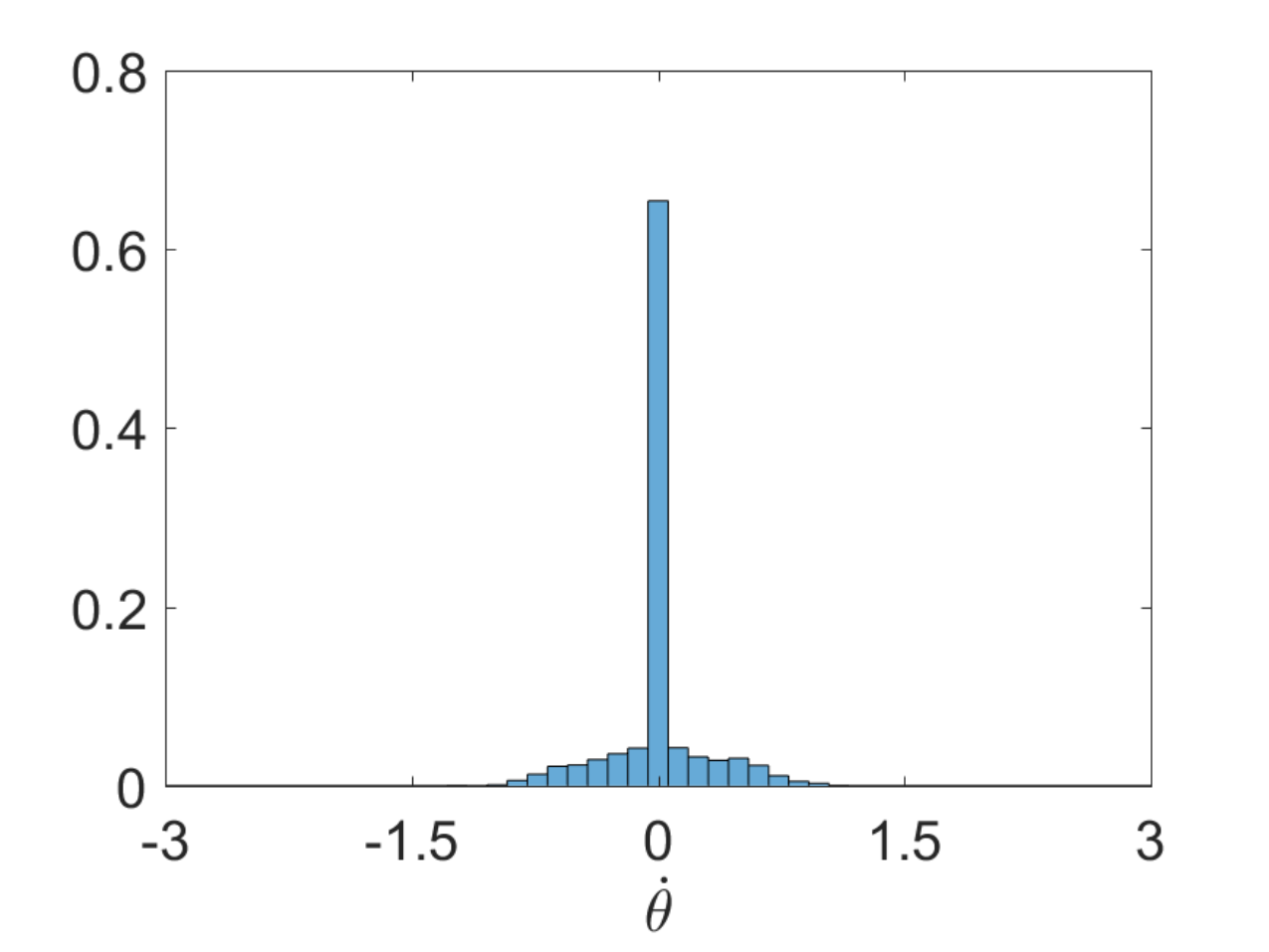}
		\end{center}
		\caption{
			\textbf{a}) A graph of an even unimodal probability density function $g$. We take a Gaussian centered at $0$ with $\sigma=0.3$.	\textbf{b}) The corresponding critical curve $\mathcal{C}$ intersects postive
			real semiaxis at a unique point $P_0$ indicated by a green star. The preimage of $P_0$ under $G$ is indicated by
			the green star in (\textbf{a}). \textbf{c}) $P_0$ corresponds to the PF bifurcation of mixing resulting in a
			PLS, which is then gradually transformed into
			synchronous state.
			\textbf{d}) The velocity distribution within the PLS
			  near PF bifurcation is determined by the eigenfunction \eqref{vPF}.
		}
		\label{f.uni}
	\end{figure}

In this section, we explain Penrose's method by applying it to the second
order model \eqref{2KM} with a unimodal density (see Fig.~\ref{f.uni}\textbf{a}).
In the next section, we apply this method to unfold a codimension-$2$ bifurcation
of mixing in the model with a bimodal density. In these two sections, we 
restrict to $W\equiv 1$, which corresponds to the all--to--all connectivity.
The bifurcation scenarios discussed below will also hold for the KM on any
graph sequence with a constant graph limit, e.g.,   Erd\H{o}s-R{\' enyi} or Paley graphs \cite{CMM18}.

For $W\equiv 1$, the only nonzero eigenvalue of $\bW$ is $\nu=1$ of multiplicity $1$. 
Thus, the equation for the eigenvalues of $\bS$ \eqref{Deqn} takes the following form
	\begin{equation}\label{GK}
		G(\lambda)=K^{-1},
	\end{equation}
	where $G$ is defined in \eqref{def-D}.

        Our goal is to locate the roots of \eqref{GK} in $\bbH_0=\{\lambda\in\mathbb{C}:~\Re \lambda >0\}$ and find a bifurcation value $K=K_c$ at which the root disappears from $\bbH_0$.
       To this end, we introduce
	$$
	G(\iu y)=\lim_{x\to 0+} G(x+\iu y)
	$$
	and denote $\mathcal{C}=\{ G(\iu t)\in\C:~t\in\R\}$.
	The Sokhotski-Plemelj formula \cite{Simon-Complex} gives
	\begin{equation}\label{SP}
		\begin{split}
			G(\iu t)& =\pi g(t)-\int_{-\infty}^\infty \frac{g(\omega)}{1+(\omega-t)^2}d\omega\\
			&+\iu\left\{ \int_0^\infty \frac{g(t+\omega)-g(t-\omega)}{\omega} d\omega
			-\int_{-\infty}^\infty \frac{(\omega-t) g(\omega)}{1+(\omega-t)^2}d\omega\right\}.
		\end{split}
	\end{equation}
	This yields the following parametric equations for $\mathcal{C}$:
	\begin{equation}\label{parC}
		\begin{split}  x&=\pi g(t) -\int_{-\infty}^\infty \frac{g(\omega)}{1+(\omega-t)^2}d\omega, \\
			y&=\int_0^\infty \frac{g(t+\omega)-g(t-\omega)}{\omega} d\omega -
			\int_{0}^\infty
			\frac{\omega \left(g(t+\omega)-g(t-\omega)\right)}{1+\omega^2}d\omega\\
			&=\int_0^\infty \frac{g(t+\omega)-g(t-\omega)}{\omega (1+\omega^2)}d\omega
		\end{split}
	\end{equation}
	for $t\in \R$.
	Note that $(x,y)\to 0$ as $t\in\pm\infty$. Thus, $\mathcal{C}$ is a bounded closed curve.

        Next, suppose $g$ is an even unimodal density (see Fig.~\ref{f.uni}\textbf{a}).
        The even symmetry of $g$ implies that $\cC$ is symmetric about the $x$--axis
        (cf.~\eqref{parC}). 
	It intersects the positive real semiaxis at a unique point $P_0=(x_0,0), x_0>0$.
	Further, note that $G(0)=x_0$ (Figure~\ref{f.uni}\textbf{a}, \textbf{b}).
	From the $x$--equation in \eqref{parC} we find
	\begin{equation}\label{x0-val}
	x_0=\pi g(0) -\int_{-\infty}^\infty \frac{g(\omega)}{1+\omega^2} d\omega
	 = \int^\infty_{-\infty} \frac{g(0) - g(\omega)}{1+\omega^2} d\omega >0.
       \end{equation}
       Define
       \begin{equation}\label{def-Kc}
         K_c:=(x_0)^{-1}.
         \end{equation}
         By the Argument Principle, the number of roots of \eqref{GK} in $\bbH_0$ is
	equal to the winding number of $\cC$ about $K^{-1}$ \cite{Penrose60}.
	Since for $K< K_c$, $K^{-1}$ lies outside $\cC$ (Fig.~\ref{f.uni}\textbf{b}), and the winding
	number is $0$. We conclude that for $K< K_c$, $\bS$ has no eigenvalues with positive real parts.
	Thus, for $K<K_c$, mixing is linearly stable\footnote{In fact, it is asymptotically
          stable with respect to a suitable weak topology \cite{ChiMed22}.}.
	For $K>K_c$, on the other hand, the winding number is $1$.
	Because of eigenvalues with positive real parts, mixing is unstable for $K>K_c$.
	As $K\to K_c+0$, $\lambda\to 0+$,  and at $K=K_c$, mixing undergoes a pitchfork (PF) bifurcation.

        Using \eqref{w-lambda} and \eqref{Ups-iy}, we compute the eigenfunction
        (written in $\omega$-variable) corresponding to $\lambda=0$:
	\begin{equation}\label{vPF}
		\Upsilon_0=\pi g(0) \delta_0+\iu\mathcal{P}_0[g]
		-\frac{e^{-\xi}}{1-\iu \omega} g(\omega).           
	\end{equation}
	The first two  terms on the right--hand side of \eqref{vPF} have singularities at $\omega =0$. The second
	term also has a regular component, that is smooth in $\omega
        \neq 0$. This determines the structure of the PLS bifurcating from
	the mixing state (Fig.~\ref{f.uni}c). The delta function on  the right--hand side of \eqref{vPF} implies that 
	the coherent cluster within the PLS is stationary. The regular component of  $\iu\mathcal{P}_0[g]$ and the third term
	yield the velocity distribution within the incoherent group. The combination of these two terms yields the velocity
	distribution within the PLS (Fig.~\ref{f.uni}\textbf{d}). {$K = 0.29$ and $n= 5000$ were used in \textbf{d}.}
\begin{figure}
		\centering
		\begin{minipage}[b]{.98\textwidth}
			\textbf{a}\includegraphics[width=0.22\textwidth]{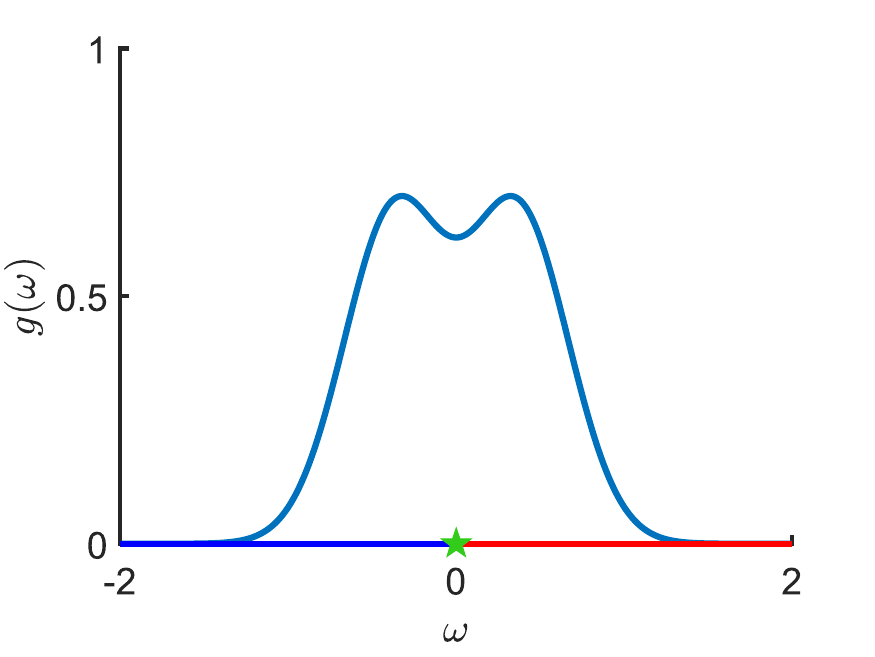}
			\hfill
			\textbf{b}\includegraphics[width=0.22\textwidth]{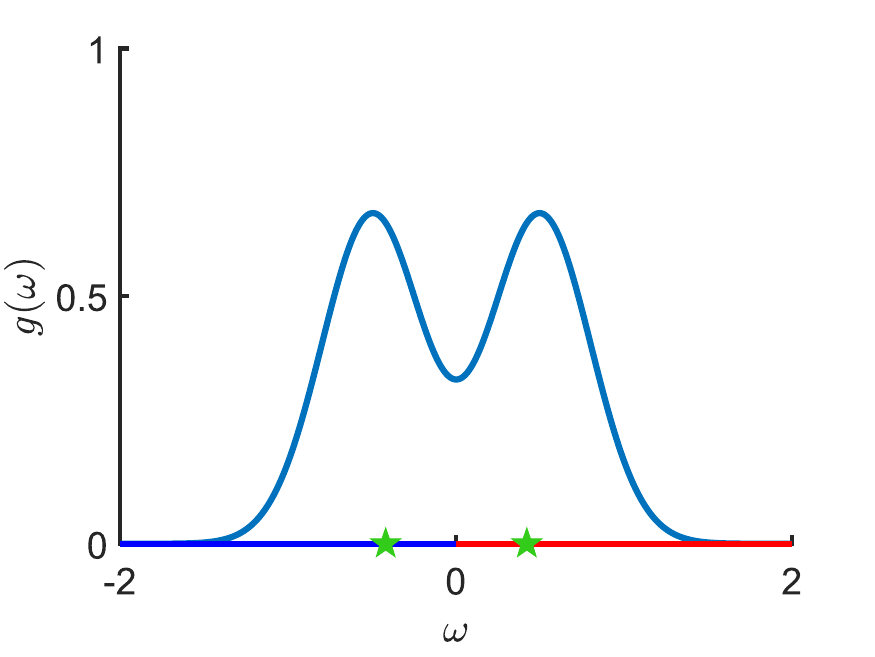}
			\hfill
			\textbf{c}\includegraphics[width=0.22\textwidth]{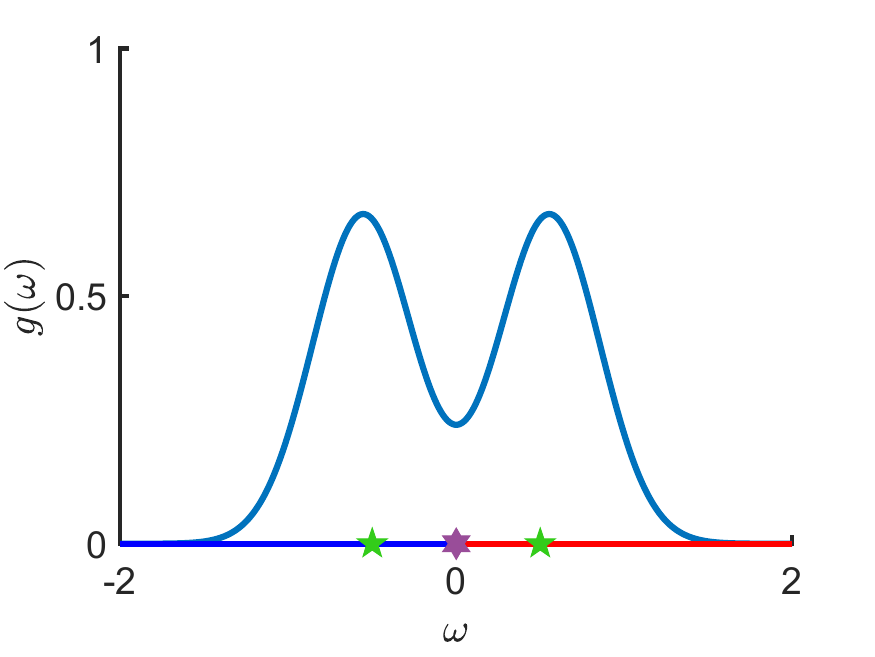}
			\hfill
			\textbf{d}\includegraphics[width=0.22\textwidth]{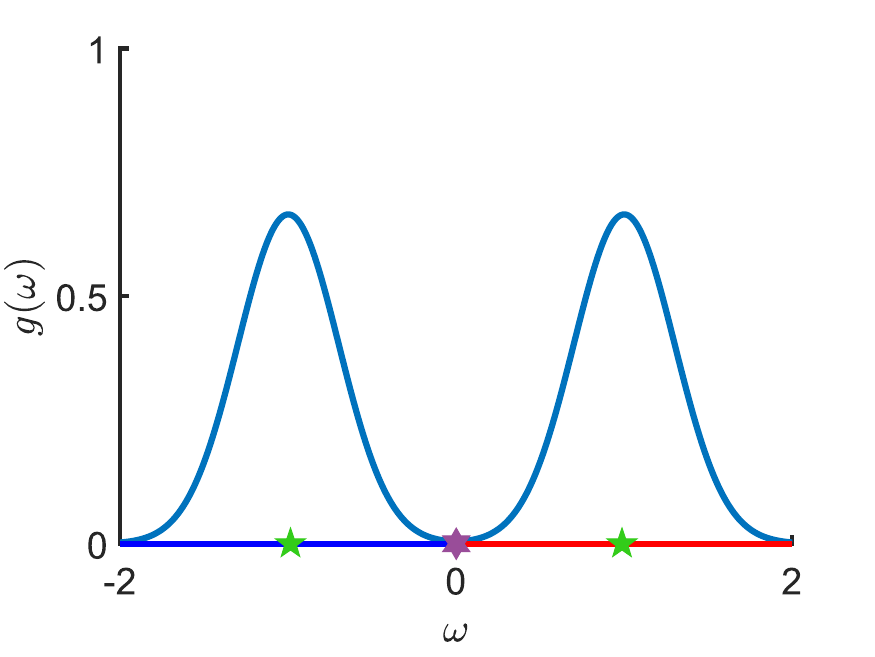}\\
			\textbf{e}\includegraphics[width=0.22\textwidth]{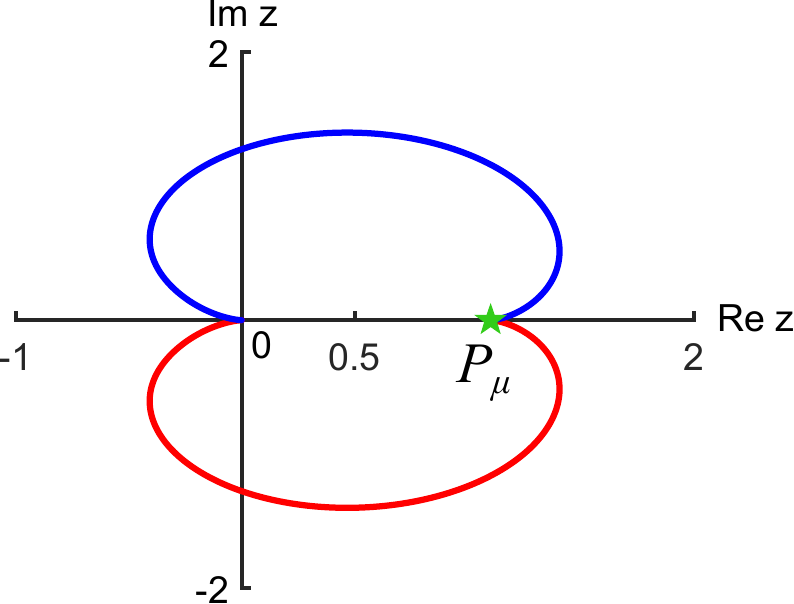}
			\hfill
			\textbf{f}\includegraphics[width=0.22\textwidth]{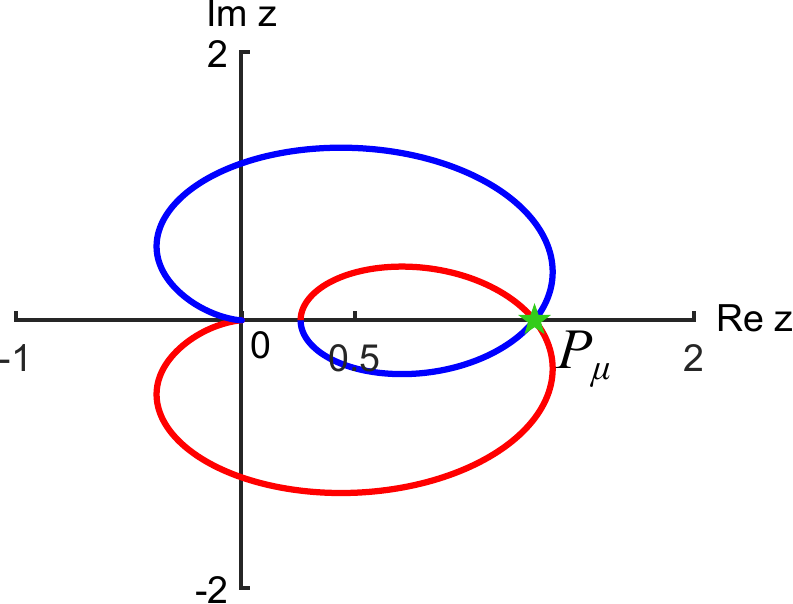}
			\hfill
			\textbf{g}\includegraphics[width=0.22\textwidth]{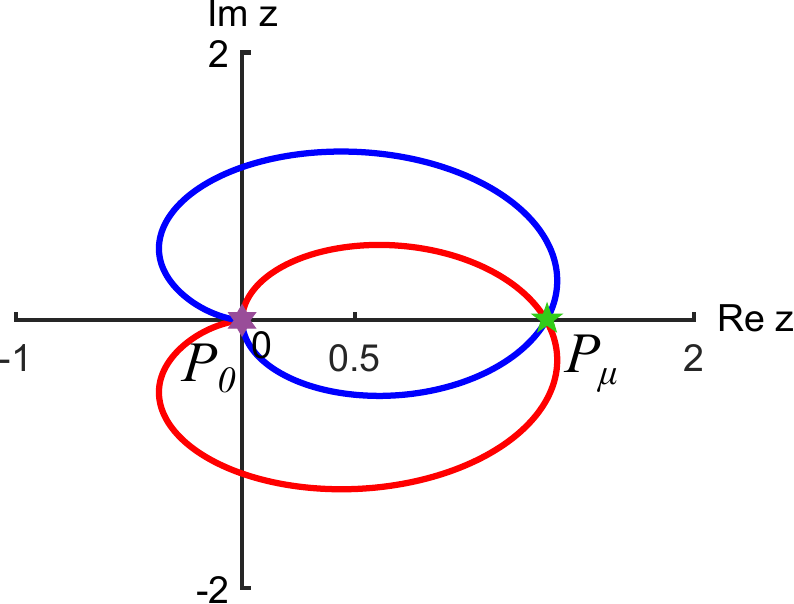}
			\hfill
			\textbf{h}\includegraphics[width=0.22\textwidth]{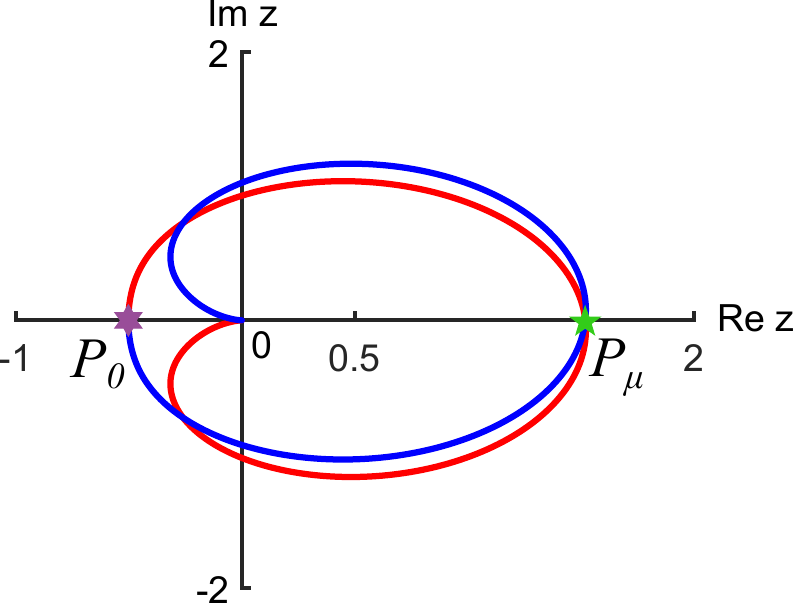}
			\hfill
			\caption{Continuous deformation of the unimodal symmetric density $g$ with $\sigma= 0.3$
				into a bimodal symmetric one (\textbf{a}-\textbf{d}) and the plots of the corresponding
				critical curves (\textbf{e}-\textbf{h}). At the critical value $\mu=\mu^\ast\approx 0.413$, $\cC_{\mu^\ast}$ develops
				a cusp (\textbf{e}). This corresponds to the codimension--$2$ bifurcation of mixing.
				The preimages of points of the intersection of the critical curve with the real
				axis $P_0$ and $P_\mu$ in (\textbf{e}-\textbf{h}) are indicated by stars in the corresponding plots in
				(\textbf{a}-\textbf{d}). From left to right, $\mu$ increases from $\mu^*$ to $0.5,$ $0.555$, and $1$.
			}\label{f.bi}
		\end{minipage}
	\end{figure}

	\section{A bimodal distribution}\label{sec.bimod}
        In this section, we study \eqref{2KM} with bimodal frequency distribution. In this setting we find
        new bifurcations of mixing: an AH bifurcation and a second PF bifurcation. They result in new
        patterns that are not present in the unimodal case. In the end of this section, we show that breaking
        symmetry in a family of bimodal distributions leads to formation of chimera states as in a similar
        scenario for the classical KM identified in our earlier work \cite{CMM20, MM22}. In the numerical
        experiments presented in this section we use the following family of probability
        density functions
	\begin{equation}\label{phi}
		g^{\mu}_{\sigma_1,\sigma_2}(x)= \frac{1}{2\sqrt{2\pi}}
		\left\{ \frac{ e^{{-(x+\mu)^2\over 2\sigma_1^2}}} {\sigma_1} +
		\frac{e^{{-(x-\mu)^2\over 2\sigma_2^2}}}{\sigma_2 }
		\right\}.
	\end{equation}
	When $\sigma_1=\sigma_2=:\sigma$, we collapse indices into one 
	$g^{\mu}_{\sigma}:=g^{\mu}_{\sigma,\sigma}$.

        \subsection{An Andronov-Hopf bifurcation}\label{sec.AH}
\begin{figure}
		\centering
		\begin{minipage}[b]{\textwidth}
			\textbf{a}\includegraphics[width=0.49\textwidth]{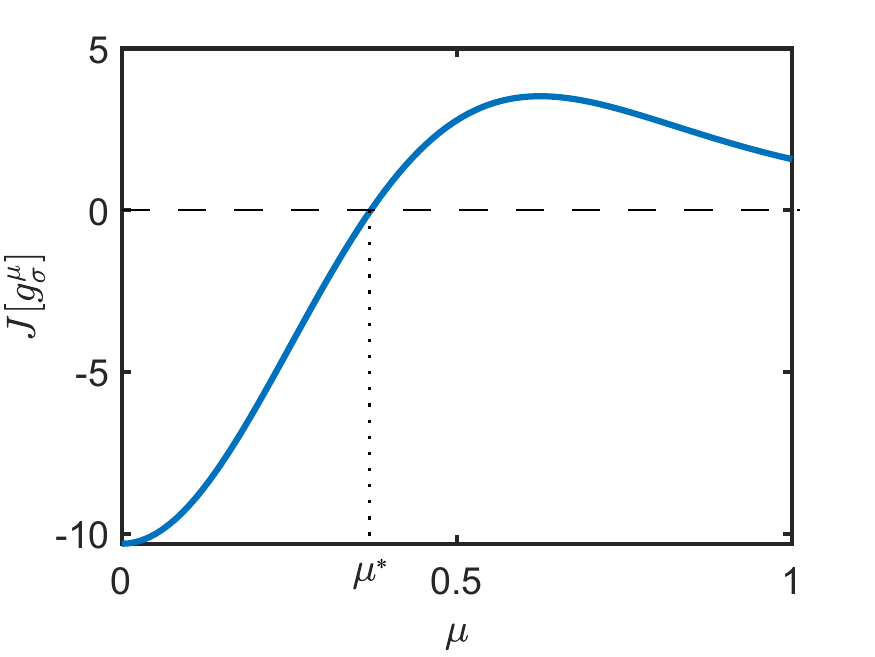}
			\textbf{b}\includegraphics[width=0.49\textwidth]{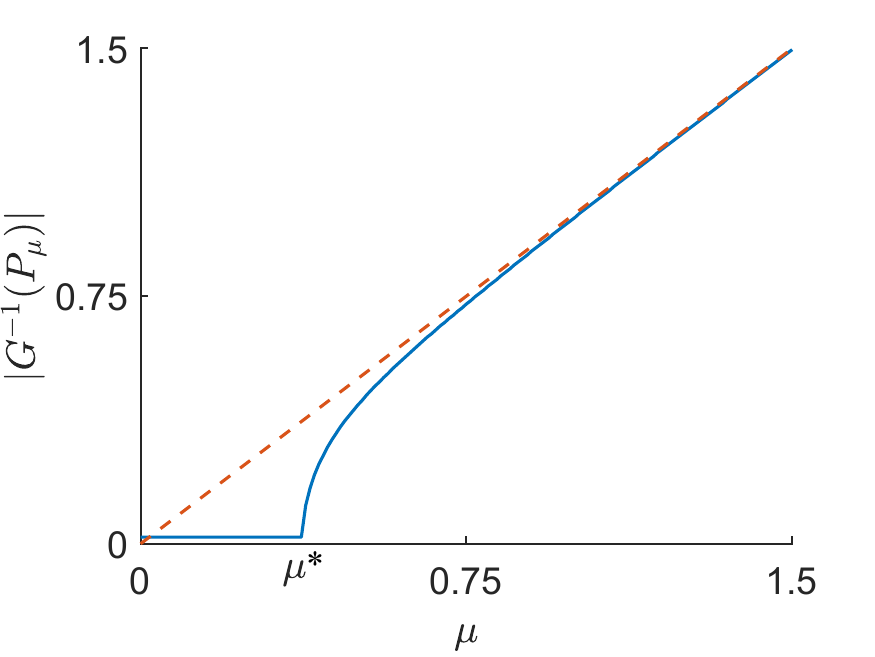}
			\caption{\textbf{a}) The plot of $J[g_\sigma^\mu]$ vs $\mu$ (cf.~\eqref{cusp}) when $\sigma = 0.3$. The zero of
				$J[g^\mu_\sigma]$ determines the critical value $\mu^\ast$.
				 \textbf{b}) The plot of the absolute value of the two preimages $G^{-1}(P_\mu)$.
				 Note that for $\mu>\mu^\ast$ outside a small neighborhood of $\mu^\ast$,
				 $\left|G^{-1}(P_\mu)\right|\approx \mu,$ i.e., the two preimages of  $P_\mu$
                                 lie near the peaks of the density $g_\sigma^\mu$.
				(cf.~\eqref{chim-mode}).} \label{f.Jplot}
		\end{minipage}
	\end{figure}

        	First, we keep $\sigma_1=\sigma_2=:\sigma$ and increase $\mu$ from
	zero.
        We want to understand how the critical curve changes as $\mu$ is varied.
	The key events in the metamorphosis of $\cC$ are shown  in Fig.~\ref{f.bi}\textbf{e}-\textbf{h}.
	For small $\mu>0$, $\cC_\mu$ \footnote{From this point on, we 
		explicitly  indicate the dependence of $\cC,$ $x,$ and $P$ on $\mu$.}
	is diffeomorphic to $\cC_0$ (cf. Fig.~\ref{f.uni}\textbf{b}) in a neighborhood
	of $P_0$, the point the intersection of $\cC_0$ with the real axis.
	At a critical value $\mu^\ast>0$,
	$\cC_{\mu^\ast}$ develops a cusp at $P_{\mu^\ast}$ (see Fig.~\ref{f.bi}\textbf{e}). To identify the condition
	for the cusp, we look for the value of $\mu$, at which the condition of the Inverse Function
	Theorem fails for $G$. By \eqref{parC} this occurs when $\left. dy/dt\right|_{t=0}=0$, i.e., 
	\begin{equation}\label{cusp}
		J[g^{\mu^\ast}_\sigma]:=
		\frac{dy}{dt}\Bigl|_{t=0} =
		2\int_0^\infty \frac{ {(g_\sigma^{\mu^\star}})^\prime(s)}{s(1+s^2)}ds=0
	\end{equation}
	(see Fig.~\ref{f.Jplot}\textbf{a}).
	
\begin{figure*}
		\centering
		\begin{minipage}[b]{.98\textwidth}
			\textbf{a}\includegraphics[width=0.48\textwidth]{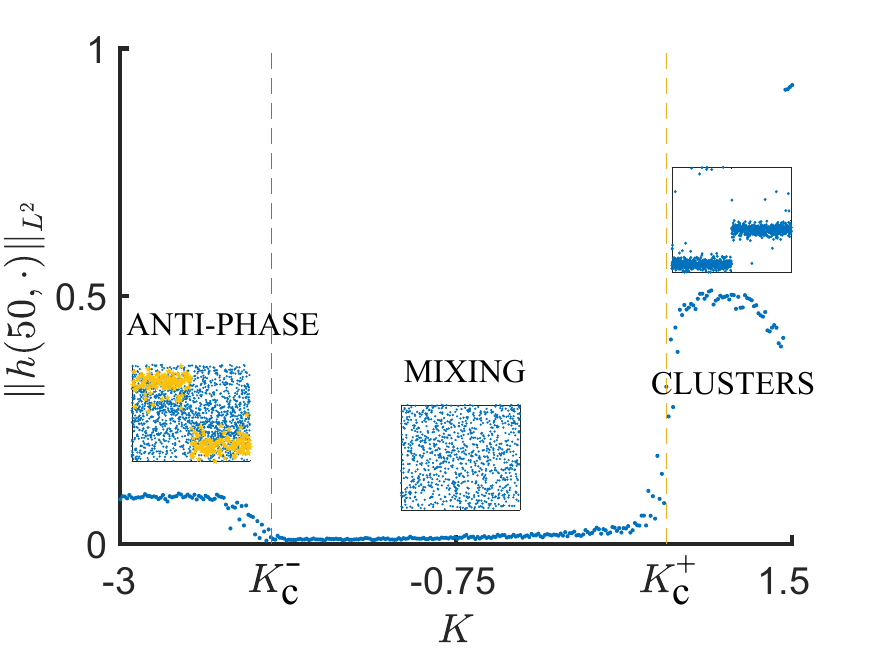}\hfill
			\textbf{b}\includegraphics[width=0.48\textwidth]{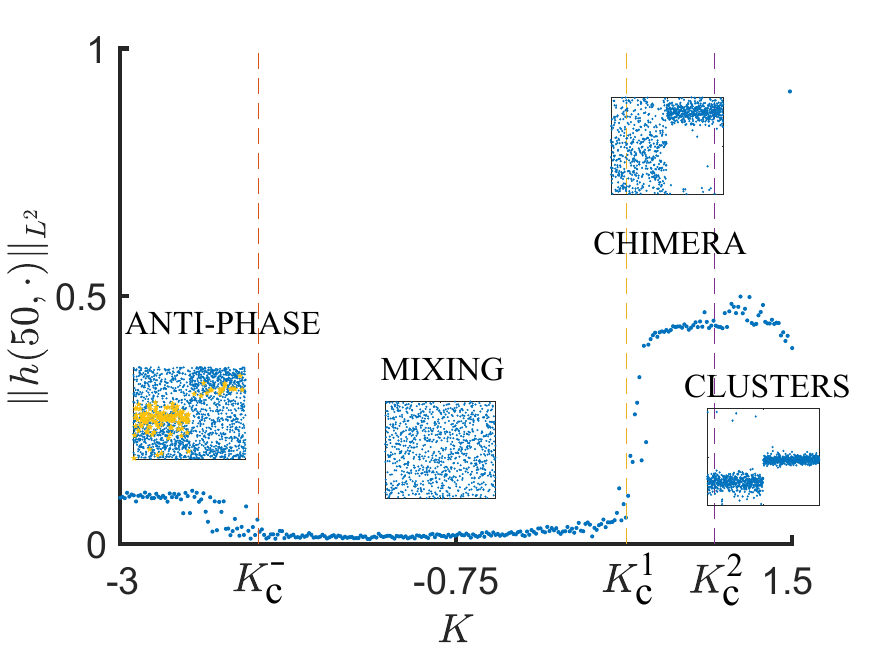}
			\caption{The bifurcation diagrams corresponding to the symmetric  and asymmetric bimodal distributions
				for all--to--all (\textbf{a} and \textbf{b}, respectively). Colored dots indicate the value of the order
				parameter computed for each cluster separately for different values of $K$ and different realizations
				of $\omega_i$'s. To improve visualization, oscillators are rearranged into two groups
                                depending on the sign of their intrinsic frequencies.
				In (\textbf{a}), the loss of stability of mixing
				at $K_c^+$ results from the AH bifurcation and so creates a traveling cluster state.
				In (\textbf{b}),
				a chimera is born at the loss of stability of mixing at $K_c^1$. It bifurcates into a moving
				traveling cluster at $K_c^2$. Note that the bifurcations at $K_c^1$ and $K_c^2$
                                affect clusters practically separately. In both cases there is an additional PF bifurcation at
                                $K_c^-<0$ resulting in stationary, anti-phase clusters. To better visualize this state,
                                all oscillators whose average velocity is sufficient small are colored in yellow stars.
                                { The following parameters were used for the distributions of the intrinsic frequencies
                                 a) $\sigma = 0.3$, $\mu = 1$; b) $\sigma_1 = 0.4$, $\sigma_2 = 0.2,$ $\mu = 1.$       }
			}\label{f.bi-bif}
		\end{minipage}
              \end{figure*}

	For $\mu>\mu^\ast$ there is a point on the real axis $P_\mu$, which has
	two preimages under $G$ denoted by $\pm \iu\nu$ (Fig.~\ref{f.bi}\textbf{b}, \textbf{f}).
	Thus, for $\mu>\mu^\ast$ mixing loses stability
	through the Andronov-Hopf (AH) bifurcation at $K = K_c^+(\mu)$, $\mu>\mu^\ast$,
        giving rise to a two-cluster pattern shown in Figure~\ref{f.patterns} \textbf{c}.
        At the AH bifurcation,
	$\bS$ has a pair of complex conjugate eigenvalues $\pm \iu \nu$. 
	The corresponding eigenfunctions written in $\omega$-variable are given by \eqref{Ups-iy}
	\begin{equation}\label{AHmode}
		\Upsilon_{\pm \iu \nu}=
		\pi g(\pm\nu) \delta_{\pm\nu}+\iu\mathcal{P}_{\pm\nu}[g]
		- \frac{e^{-\xi}}{1+\iu (\pm\nu-\omega)} g(\omega).
              \end{equation}
              Tempered distributions $\Upsilon_{\iu \nu}$ and  $\Upsilon_{-\iu \nu}$ have singularities at $\iu\nu$ and $-\iu\nu$
              respectively due to $\delta_\cdot$ and $\mathcal{P}_\cdot$ on the right--side of \eqref{AHmode}.
              This implies the existence of two groups of phase-locked oscillators moving with velocities approximately
              equal to $\pm \nu$. Moreover, Fig.~\ref{f.Jplot}{\bf b} shows that outside a
              small neighborhood of $\mu^*$, $\nu \approx \mu$ and so the group velocities
              correspond to the peaks of the density $g_\sigma^\mu$. The regular
              part of $\mathcal{P}_\cdot$ results in a cloud of irregularly moving oscillators.
            This explains the salient features of the two-clusters patterns in the pattern replacing mixing
            after it loses stability (see Fig.~\ref{f.bi-bif}\textbf{a}).
            Note that $\mu=\mu^*$ separates the regions of the PF and AH bifurcations. At this value of $\mu$
            and the corresponding critical value of $K$ mixing undergoes a codim-2 bifurcation. Unfolding of this
            bifurcation contains a range of spatiotemporal patterns bifurcating from mixing including one- and
            (traveling) two- cluster states and chimera states (see Section~\ref{sec.chimeras}).

	\subsection{ A second pitchfork bifurcation}
	For increasing values of $\mu>\mu^\ast$, the loop formed by the
	critical curve grows while remaining in the right half-plane (Fig.~\ref{f.bi}\textbf{f}). At a certain value
	$\mu^0>\mu^\ast$ it hits the origin (Fig.~\ref{f.bi}\textbf{g}).
	For $\mu>\mu^0$ the point of simple intersection of $\mathcal{C}$ with the real axis moves into
	the negative semiaxis. This corresponds to the creation of the new pitchfork bifurcation at a negative
	value:
	\begin{equation}\label{neg-K}
		K_{c}^-=\left(\pi g(0) -\int_{-\infty}^\infty \frac{g(\omega)}{1+\omega^2} d\omega\right)^{-1}<0,
              \end{equation}
              which leads to a pattern shown in Figure~\ref{f.patterns} \textbf{g}.
              Thus, for $\mu>\mu^0$ mixing is stable for $K\in (K_{c}^-, K_{c}^+)$ with $K_{c}^-<0<K_{c}^+$.
              The corresponding unstable mode at the PF bifurcation at $K_{c}^-$ is still given by \eqref{vPF} albeit with a
              bimodal $g$. In the present case, Equation \eqref{vPF} implies that there is a group of stationary phase-locked oscillators
              due to $\delta_0$ and the singularity of $\mathcal{P}_0[g]$ on the right--hand side of \eqref{vPF}.
              In addition, there is a group of moving oscillators
              whose velocities are determined by the regular part of  $\mathcal{P}_0[g]$ and the last term
              on the right--hand side of \eqref{vPF}.

              Equation \eqref{vPF} accounts for the velocity distribution of the pattern replacing mixing but it
              does not explain why the phase-locked oscillators are organized in two antiphase coherent
              groups
              whereas for the PF at positive $K_{PF}$ analyzed in Section 3 there is a single coherent group.
              The splitting into two groups can be understood with the help of the method used in
              \cite{MM22} for studying cluster
              dynamics. We outline the argument from \cite{MM22} to the extent needed for present
              purposes. To this end, let
              $$
              J^-:=\{j\in [n]:\; \omega_j<0\}\quad\mbox{and}\quad J^+:=\{j\in [n]:\; \omega_j\ge 0\}
              $$
              and
              $$
              U_1=|J^-|^{-1}\sum_{i\in J^-} \theta_i \quad\mbox{and}\quad  U_2=|J^+|^{-1}\sum_{i\in J^+} \theta_i.
              $$
              Here, $|J|$ denotes the cardinality of $J$.  $U_{1,2}$ describe the evolution of the
              two macroscopic clusters of phase-locked oscillators. In \cite{MM21} it is shown that in the limit $n\to\infty$,
              $U_1$ and $U_2$ satisfy the following system of ODEs
              \begin{equation*}
                \begin{split}
                  \ddot U_1+\dot U_1 &=-\mu + K\sin\left(U_2-U_1\right),\\
                  \ddot U_2+\dot U_2 &= \mu + K\sin\left(U_1-U_2\right),
                \end{split}
              \end{equation*}
              From this, we derive an ODE for $X=U_2-U_1$
              \begin{equation}\label{eqn-X}
                \begin{split}
                  \dot X &= Y,\\
                  \dot Y &= 2\mu -Y-K \sin X.
                  \end{split}
              \end{equation}
              A standard calculation shows that $(\arcsin\left(\frac{2\mu}{K},0\right)$ and
                $\left(\pi-\arcsin\left(\frac{2\mu}{K} \right), 0\right)$ are two equilibria of
                \eqref{eqn-X}. Furthermore, the former is stable for $K>0$ and the latter
                is stable for $K<0$.
                      
              \begin{figure}
		\centering
		\begin{minipage}[b]{.98\textwidth}
			\textbf{a}\includegraphics[width=0.29\textwidth]{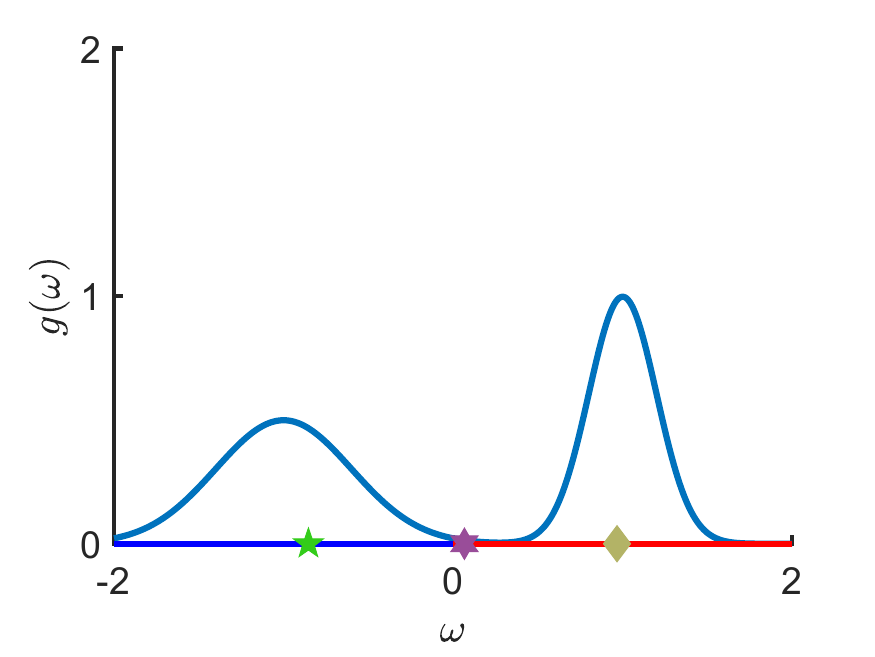}
			\hfill
			\textbf{b}\includegraphics[width=0.29\textwidth]{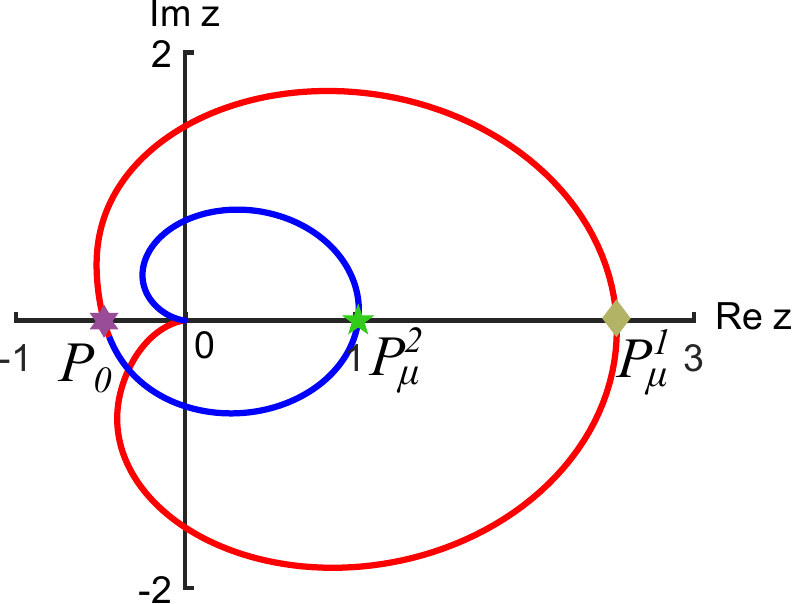}
                        \textbf{c}\includegraphics[width=0.29\textwidth]{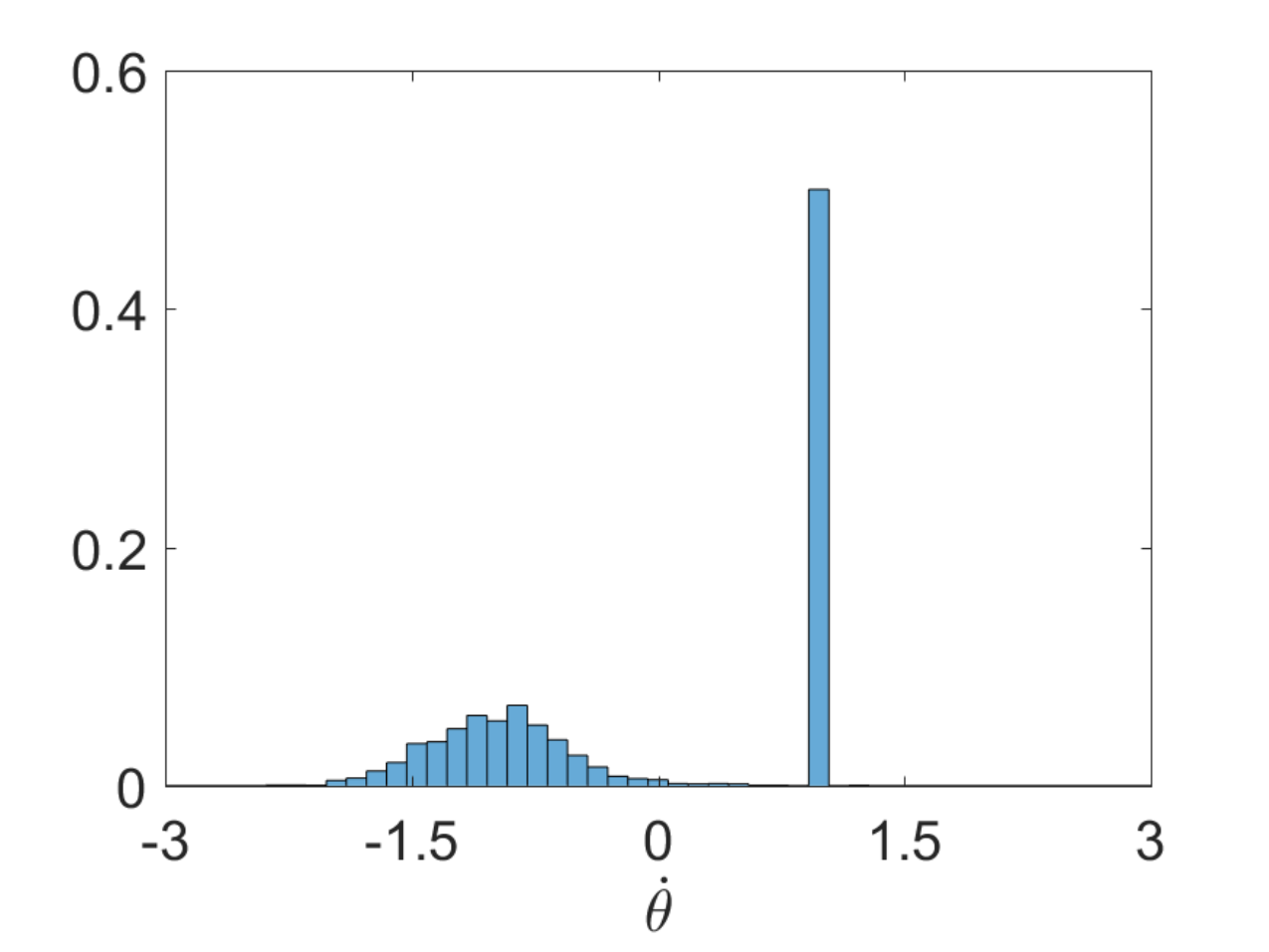}
			\caption{ {\bf a}) Asymmetric bimodal distribution with $\sigma_1 = 0.4$,
                          $\sigma_2 = 0.2$, and $\mu = 1$; {\bf b})
                          its corresponding critical curve.
                          \textbf{c}) The histogram of the
				velocity distribution
				within a chimera is fully determined by the singular  distribution $v_{\iu t_\mu+0}$.
                               {
                                The histogram
                                was generated by simulating \eqref{2KM} with $n=5000, K = 0.75$.
                                }
			}\label{f.asym}
		\end{minipage}
	\end{figure}

        \subsection{Chimera states}\label{sec.chimeras}
	We now fix $\mu\in (\mu^\ast, \mu^0)$ and break the even symmetry of $g^\mu_\sigma$ by
	increasing $\sigma_1$
	 and decreasing $\sigma_2$
	(see Fig.~\ref{f.asym}\textbf{a}).
	This affects the critical curve $\cC_{\mu,\sigma_1,\sigma_2}$ in the following way.
	The point of double intersection $P_\mu$ splits into two points of intersection with the
	real axis: $P_\mu^1=(x_\mu^1, 0)$ and $P_\mu^2=(x_\mu^2, 0)$ with $0<x_\mu^2<x_\mu^1$
	(see Fig.~\ref{f.asym}\textbf{b}).
	Note that the preimages of these points under $G$ are still very close to the maxima of
	$g^\mu_{\sigma_1,\sigma_2}$ (see Fig.~\ref{f.asym}\textbf{a}).
	In particular, the preimage of $P_\mu^1$
	is approximately $\iu\mu$, the center of the more localized peak of $g^\mu_{\sigma_1,\sigma_2}$ .
	This implies that mixing loses stability at
        $ K_c^1\approx G(\iu\mu)^{-1}$
	The bifurcating eigenvalue
	$\lambda=\iu\nu_1 (\nu_1\approx \mu)$ and the corresponding eigenfunction
	\begin{equation}\label{chim-mode}
		\Upsilon_{\iu\nu_1} = \pi g^{\mu}_{\sigma_1,\sigma_2}(\nu_1) \delta_{\nu_1} +
		\iu \mathcal{P}_{\nu_1}[g^{\mu}_{\sigma_1,\sigma_2}]
		- \frac{e^{-\xi} g_{\sigma_1,\sigma_2}^\mu (\omega)}{1+\iu (\nu_1-\omega)}.
	\end{equation}
	Note that the first term on the right hand side of \eqref{chim-mode} is a singular distribution
	localized at $\nu_1$. The second term  has a singularity at $\nu_1$, but its regular part  has some
	`weight' near $\nu_2\approx -\mu$. These features translate into the velocity distribution within a chimera:
	there is a tightly localized peak around $\mu$ (the coherent group) and a broader
	peak near $-\mu$ (the incoherent group) (Figs.~\ref{f.asym}\textbf{c} and \ref{f.bi-bif}\textbf{b}).

	\begin{figure*}
		\centering
		\begin{minipage}[b]{.99\textwidth}
			\textbf{a}\includegraphics[width=0.32\textwidth]{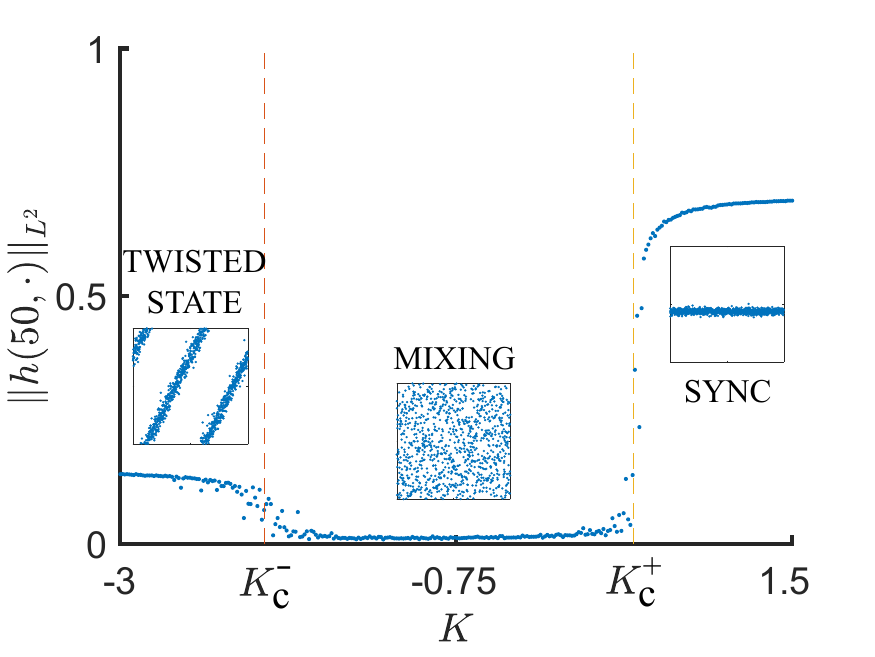}\hfill
			\textbf{b}\includegraphics[width=0.32\textwidth]{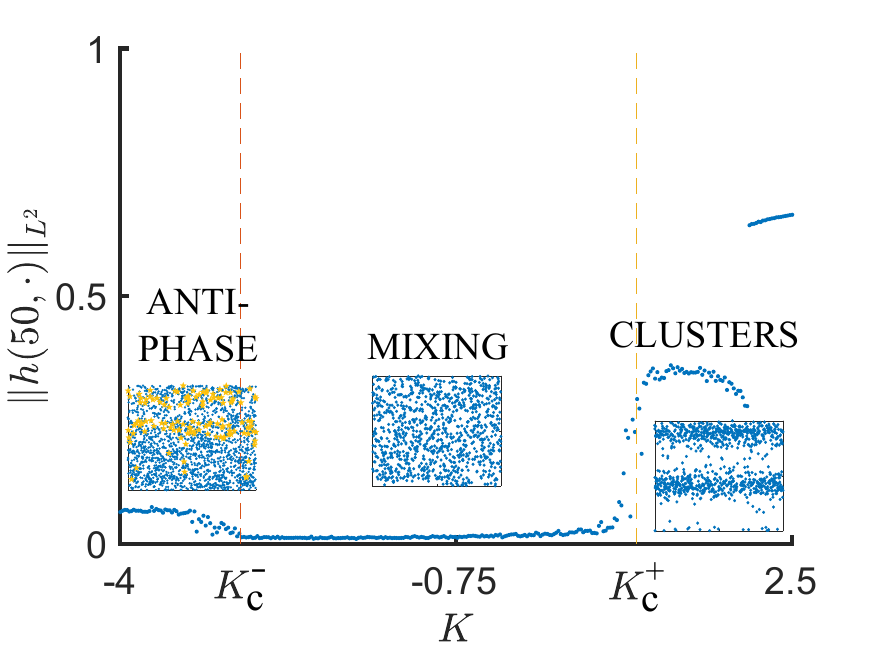}\hfill
			\textbf{c}\includegraphics[width=0.32\textwidth]{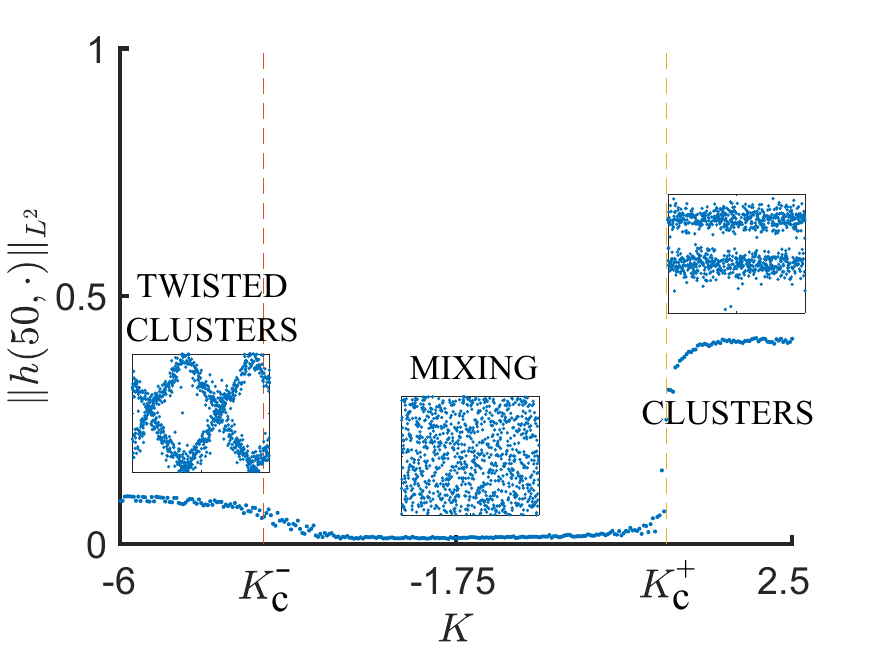}
			\caption{Bifurcation diagrams for \eqref{2KM} with unimodal and bimodal
                          intrinsic frequency distributions and nearest neighbor coupling with range $r = 0.35$.
                          For this coupling type, $\mathbf{W}$ has eigenvalues of both signs, i.e., 
                          $\nu^-<0<\nu^+$. {\bf (a)} For the unimodal distribution ($\sigma = 0.3$)
                          the critical curve has a (simple) intersection with the real axis at $x^+>0$ resulting in bifurcations
                          $K_c^- = -\frac{1}{|\nu^- x^+|}$ (stationary twisted states) and $K_c^+ = \frac{1}{\nu^+ x^+}$
                          (synchronization). {\bf (b)} For the bimodal distribution $(\sigma_1=\sigma_2=0.3$, $\mu = 1$) the
                          critical curve has two intersections with the real axis, $x^- < 0 < x^+$ (where $x^-$ results in
                          a PF and $x^+$ in an AH bifurcation). In this case $K_c^- = -\frac{1}{|\nu^+ x^-|}$
                          (stationary, anti-phase clusters) and $K_c^+ = \frac{1}{\nu^+ x^+}$
                          (moving homogeneous clusters). {\bf (c)} Varying the bimodal distribution
                          $(\sigma_1 = \sigma_2 = 0.3$, $\mu = 2$) changes the roots of the critical curve.
                          In this case $K_c^- = -\frac{1}{|\nu^-x^+|}$ (moving twisted clusters) and
                          $K_c^+ = \frac{1}{\nu^+x^+}$ (moving homogeneous clusters).
			}\label{f.nn-bif}
		\end{minipage}
	\end{figure*}

	\section{ The role of connectivity: nearest--neighbor coupling}\label{sec.connect}
        \setcounter{equation}{0}

        We have seen above that a PF bifurcation of mixing results in the formation of one or a pair of
        stationary coherent clusters depending on the distribution type and the sign of $K$, while the AH bifurcation
        leads to a pair of travelling coherent structures. All these patterns are spatially homogeneous if the coupling
        is all-to-all, because the only nonzero eigenvalue of $\bW$ is
        positive and the corresponding eigenfunction is constant.
         In general, $\bW$
	 may have  eigenvalues of both signs \cite{ChiMed19a}. In this case, the interval of stability of mixing
         is bounded from both sides $(K_c^-,K_c^+)$, $K_c^-<0<K_c^+$. The values of $K_c^-$ and $K_c^+$,
         as well as the types of the bifurcations at these values of $K$, depend on the interplay between the type of the
         distribution of $\omega_i$ and the spectral properties of $\bW$. Furthermore, the bifurcation at one
         of these points results in a pattern with nontrivial spatial structure. In this section, we illustrate some
         of possible bifurcation scenarios by considering \eqref{2KM} with a nonlocal nearest-neighbor coupling.

	Let $W(x,y)=U(x-y)$, which is defined by
	$$
	U(x)=\1_{(-r,r)}(x), \quad \mbox{on} \; (-1/2, 1/2)
	$$
	and extended to $\R$ by periodicity. Here, $\1_{\mathbf{A}}$ stands for the indicator function,
	and $r\in (0,1/2)$ is a fixed parameter. Then
	$$
	\bW[f](x)=\int_{-1/2}^{1/2} U(x-y) f(y)dy.
	$$
	The eigenvalues of $\bW$ can be computed explicitly
	$$
	\nu_k=\int_{-1/2}^{1/2} U(x) e^{\pm 2\pi\iu kx}dx= \int_{-1/2}^{1/2} U(x) \cos\left(2\pi kx\right) dx, \quad k=0,1,2,\dots.
	$$
	The corresponding eigenfunctions are $w_k=e^{\pm 2\pi\iu kx}$.
	The largest positive eigenvalue is $\nu^+ := \nu_0 = 2r$ (cf.~\cite[Lemma~5.3]{ChiMed19a}).
	Since $k=0$, the corresponding eigenspace is $1$-dimensional consisting of constant functions.
	By $k^\ast>0$ denote the value of $k$ corresponding to  the smallest negative eigenvalue
	of $\bW$, $\nu^- := \nu_{k^\ast}$. The corresponding eigenfunctions are $e^{2\pi\iu k^\ast x}$ and
	$e^{-2\pi\iu k^\ast x}$.
	
	To explain the implications of the presence of the eigenvalues of both signs in the spectrum
	of $\bW$, we first turn to the unimodal distribution. If $g$ is even and unimodal  then the region
	of stability of mixing is a bounded interval $(K_c^-, K_c^+)$ with $K_c^-=(\pi g(0)\nu^-)^{-1}$
	and $K_c^+=(\pi g(0)\nu^+)^{-1}$ \cite{ChiMed19a}. At $K_c^+$ we observe a familiar scenario
	of transition to synchronization (Figure~\ref{f.nn-bif}\textbf{a}). At $K_c^-$ the situation is different.
	The center subspace 
	of the linearized problem in the Fourier space is spanned by
	$$
	v_{\nu^-}^{(1)} = \Upsilon_0(\omega, \xi) e^{2\pi\iu k^\ast x} \quad\mbox{and}\quad
	v_{\nu^-}^{(2)} = \Upsilon_0(\omega, \xi) e^{-2\pi\iu k^\ast x}.
	$$
	In the solution space, we therefore expect that
	\begin{equation}\label{invert}
          \begin{split}
	f(t,\theta,\psi,\omega,x)&\sim
	\operatorname{Re} \left\{\int_\R e^{\iu(\theta-\zeta)} \left( c_1 v_{\nu^-}^{(1)}+c_2 v_{\nu^-}^{(2)}\right) d\zeta,\right\}\\
        &=\operatorname{Re} \left\{ \left(c_1 e^{\iu (2\pi k^\ast x+\theta)}+  c_2 e^{\iu (-2\pi k^\ast x+\theta)}\right)
          \tilde\Upsilon_0(\omega, \psi)\right\}, \quad 	c_1,c_2\in\C,
      \end{split}
    \end{equation}
    and
    $$
    \tilde\Upsilon_0(\omega, \psi)=\int_\R e^{-\iu\zeta\psi} \Upsilon_0\left(\omega, \xi(\zeta) \right)d\zeta.
    $$
    For the PLS emerging at the bifurcation, we see that the structure encoded in $\tilde\Upsilon_0(\omega, \psi)$
	is now  superimposed onto a linear combination of $\pm k^\ast$--twisted states
	(Fig.~\ref{f.nn-bif}\textbf{a}).
	
	The same principle applies to the analysis of bifurcations in the bimodal case. Suppose $\mu$ and
	$\sigma$ are such that the critical curve has the form as shown in Figure~\ref{f.bi}\textbf{h}.
	Recall that $x^-<0<x^+$ denote the $x$-coordinates of the points of intersection of the critical
	curve with real axis, $P_0$ and $P_\mu$. The former is a simple intersection point and the latter is a
	double intersection point. The expression for $x^-$ is known explicitly
	\be\lbl{x-}
	x^-=\pi g(0) -\int_{-\infty}^\infty \frac{g(\omega)}{1+\omega^2} d\omega.
	\ee
	As in the unimodal case, mixing is stable in a finite interval for $K$, $(K^-_c, K^+_c)$. The values
	of $K_c^-$ and $K_c^+$ as well as the types of the bifurcations at these points depend
	on $x^-$, $x^+$, $\nu^-$, and $\nu^+$:
	\be\lbl{K-bi}
	K_c^-=-\min\left\{ \frac{1}{|\nu^+x^-|}, \frac{1}{|\nu^-x^+|} \right\}\quad\mbox{and}\quad
	K_c^+=\min\left\{ \frac{1}{\nu^+x^+}, \frac{1}{\nu^-x^-} \right\}.
	\ee
	Here, the type of the intersection at $x^\pm$ (simple vs double) determines the type of the bifurcation,
	while the eigenfunctions $V^\pm$ corresponding to $\nu^\pm$ determine the spatial organization
	of the emerging pattern (homogeneous vs twisted states). Note that each of the two possible values
        of $K_c^-$ and $K_c^+$ in \eqref{K-bi} corresponds to a distinct combination of the velocity distribution
        and the spatial profile of the emerging pattern. This results in a four distinct bifurcation scenarios for the
        loss of stability of mixing in the KM with symmetric bimodal intrinsic frequency distribution.
	
	To illustrate different bifurcation scenarios, we use the following examples.
	Suppose $|\nu^+x^-| >|\nu^-x^+|$ then $K_c^-=(\nu^+ x^-)^{-1}$. Because $x^-$ is a simple intersection
	point, the corresponding bifurcation is PF. The anti-phase solution bifurcating from mixing at $K=K_c^-$ is
	shown in Fig.~\ref{f.nn-bif}\textbf{b} (compare with the anti-phase solution in Fig.~\ref{f.bi-bif}\textbf{b}).
	Alternatively, if $|\nu^+x^-| <|\nu^-x^+|$ then $K_c^-=(\nu^- x^+)^{-1}$. This time the bifurcation is AH
	and the bifurcating pattern are two sets of traveling twisted states (Fig.~\ref{f.nn-bif}\textbf{c}).
	
	Likewise, there are two possible scenarios for the bifurcations at $K=K_c^+$. If $\nu^+x^+ >\nu^-x^-$ then
	$K_c^+=(\nu^+ x^+)^{-1}$. Thus, we have an AH bifurcation producing two sets of traveling clusters (Fig.~\ref{f.nn-bif}\textbf{c}).
	If, on the other hand, $\nu^+x^+ <\nu^-x^-$ then $K_c^+=(\nu^- x^-)^{-1}$. The corresponding bifurcation
	is PF producing a set of stationary twisted states. To illustrate the last scenario, we take
        $$
        U(x) = 2\nu \cos(2\pi x).
        $$
        With this choice  of $U$,  the only nonzero eigenvalue of $\bW$ is $\nu$. Taking $\nu<0$
        we have $\nu^- = \nu$ and $\nu^+$ does not exist.
    
\begin{figure}[h]
		\centering
		\includegraphics[width=0.48\textwidth]{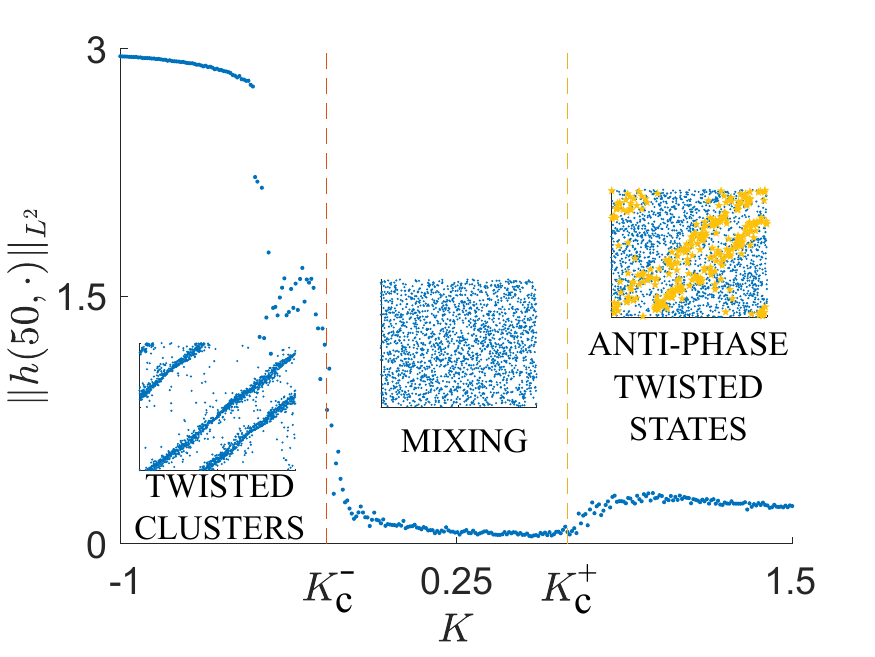}
		\caption{Bifurcation diagram for the coupling $U(x) = -6 \cos(2\pi x)$.
                    Taking bimodal distribution $(\sigma_1=\sigma_2= 0.3$, $\mu = 1$)
                    results in the bifurcations $K_c^- = -\frac{1}{|\nu^- x^+|}$ (moving twisted  clusters) and
                    $K_c^+ = \frac{1}{\nu^- x^-}$ (stationary, anti-phase twisted states). To better visualize this
                    latter state, the position of every oscillator whose average velocity is sufficiently small is
                   depicted by a yellow star.
		}\label{f.cosine}
	\end{figure}

       The bifurcation at $K_c^- = -\frac{1}{|\nu^- x^+|}$ generates moving twisted clusters, while that at 
       $K_c^+ = \frac{1}{\nu^- x^-}$ results in a pair of stationary antiphase twisted states.
       These  patterns are shown in Figure~\ref{f.cosine}.
	
 \section{Discussion}\label{sec.discuss}     
\setcounter{equation}{0}
The instability of mixing in the original KM and in the model with inertia reveals a wealth of spatiotemporal
patterns in these models. In our previous work \cite{CMM20, MM22}, we developed a method for studying these
patterns, which is based on the combination of the linear stability analysis of mixing (cf.~\cite{ChiMed19a}) and
Penrose diagrams \cite{Penrose60} (see also \cite{Die16}). In the present paper, we extend this approach to the
KM with inertia. We show that in addition to a PF and an AH bifurcations of mixing similar to those analyzed
in \cite{CMM20, MM22}, the KM with inertia features new bifurcation scenarios which were not present in 
the original model in similar settings. In particular, for the model with symmetric bimodal frequency distribution
we identify a new PF bifurcation which follows the AH bifurcation of mixing. The new PF bifurcation results in
a new PLS, which consists of two stationary clusters superimposed onto a cloud of irregularly moving oscillators
(Fig.~\ref{f.patterns}~\textbf{g}).
Note that in the original KM clusters are born in an AH bifurcation and are automatically traveling (cf.~\cite{MM22}).
The same
bifurcation in the model with inertia on nonlocal nearest neighbor graphs produces similar patterns with coherent
clusters organized as twisted states (Fig.\ref{f.patterns}~\textbf{f}). These patterns were not present in the
analysis of the original KM. Furthermore, the presence of the second PF bifurcation enriches the repertoire of possible
bifurcation scenarios considerably. For instance, in the model with a family of symmetric bimodal distributions
we find four distinct bifurcation scenarios of mixing (cf.~Section~\ref{sec.connect}) versus a single bifurcation scenario
found for the original KM in a similar setting. This underscores the flexibility of pattern forming mechanisms
in the model with inertia.

In addition to applications to biological systems well-known for the ordinary KM, the model with inertia is also
known for its applications in modeling power grids \cite{DB12}. In particular, the model with bimodal
frequency distribution comes up in the context of certain  high-voltage power grids (cf.~\cite{TOS19}).
Therefore, bifurcations of mixing identified in the present paper may be useful for understanding 
stability of these technological systems.

\vskip 0.2cm
\noindent
{\bf Acknowledgements.} This work was supported by
JST Moonshot Research and Development grant No. JPMJMS2023 (to HC),
NSF grant DMS 2009233 (to GSM), and by
Support of Scholarly Activities Grant at The College of New Jersey (to MSM).
Numerical simulations were completed using the high performance computing
cluster (ELSA) at the School of Science, The College of New Jersey. Funding of ELSA is provided
in part by NSF OAC-1828163.

\noindent\textbf{Data availability statement.} Data sharing is not
applicable to this article as no datasets were generated or analysed
during the current study.


\def\cprime{$'$} \def\cprime{$'$}
\providecommand{\bysame}{\leavevmode\hbox to3em{\hrulefill}\thinspace}
\providecommand{\MR}{\relax\ifhmode\unskip\space\fi MR }
\providecommand{\MRhref}[2]{%
  \href{http://www.ams.org/mathscinet-getitem?mr=#1}{#2}
}
\providecommand{\href}[2]{#2}

\end{document}